\documentclass[final,3p,times,twocolumn]{elsarticle}

\usepackage{graphicx} 
\usepackage{rotating}
\usepackage{supertabular,booktabs}
\usepackage{lscape}
\usepackage[center]{caption}
\usepackage{capt-of}
\usepackage[english]{babel}
\usepackage{float}
\usepackage{hyperref}
\usepackage{array}
\usepackage{algorithmic}
\usepackage{textcomp}
\usepackage{longtable}
\usepackage{tabularx}
\usepackage{pdflscape}
\usepackage{amsmath}
\usepackage{enumitem}
\usepackage{hyperref}
\usepackage[table]{xcolor}
\usepackage{amsmath}
\usepackage{tikz}
\usetikzlibrary{tikzmark}
\usetikzlibrary{arrows.meta}

\usepackage{amssymb}
\usepackage{pifont}
\usepackage{multirow}
\usepackage{tabularx}
\hypersetup{colorlinks=true,linkcolor=blue,	citecolor=blue,}
\usepackage{dblfloatfix}
\usepackage{footnote}
\usepackage{makecell}
\usepackage{enumitem}
\usepackage{mathtools}
\usepackage{tikz}
\usepackage{tabularx}
\renewcommand{\arraystretch}{1.12}
\newcommand*\emptycirc[1][1ex]{\tikz\draw (0,0) circle (#1);} 
\newcommand*\fullcirc[1][1ex]{\tikz\fill (0,0) circle (#1);} 
\newcommand*\halfcirc[1][1ex]{%
\begin{tikzpicture}
       \draw[fill] (0,0)-- (90:#1) arc (90:270:#1) -- cycle ;
       \draw (0,0) circle (#1);
\end{tikzpicture}}
\usetikzlibrary{tikzmark}
\usetikzlibrary{arrows.meta}
\usetikzlibrary{tikzmark}
\usetikzlibrary{arrows.meta}
\usepackage{rotating}
\usepackage[english]{babel}
\usepackage{parskip}
\usetikzlibrary{arrows.meta}
\usepackage{array}
\usepackage[Symbol]{upgreek}
\usepackage{booktabs}
\usepackage{xcolor}
\def\BibTeX{{\rm B\kern-.05em{\sc i\kern-.025em b}\kern-.08em T\kern-.1667em\lower.7ex\hbox{E}\kern-.125emX}}
\usepackage{pifont}
\usepackage{algorithmic}
\usepackage{textcomp}
\usepackage{forest,array}
\usepackage[utf8]{inputenc}
\usepackage{tabularray}
\usepackage{lscape}
\usepackage{titlesec}

\journal{Computer \& Security}

\begin{document}

\begin{frontmatter}

\title{Privacy Preservation Techniques (PPTs) in IoT Systems: A Scoping Review and Future Directions}


\author{Emmanuel Dare Alalade, Ashraf Matrawy} 

\affiliation{organization={School of Information Technology},
            addressline={Carleton University}, 
            city={Ottawa},
            postcode={K1S 5B6}, 
            state={Ontario},
            country={Canada}}





\begin{abstract}
Privacy preservation in Internet of Things (IoT) systems requires the use of privacy-enhancing technologies (PETs) built from innovative technologies such as cryptography and artificial intelligence (AI) to create techniques called privacy preservation techniques (PPTs). These PPTs achieve various privacy goals and address different privacy concerns by mitigating potential privacy threats within IoT systems. This study carried out a scoping review of different types of PPTs used in previous research works on IoT systems between 2010 and 2025 to further explore the advantages of PPT methods in these systems.
This scoping review looks at PPT methods to achieve PPT goals, possible technologies used for building PET, the integration of PPTs into the computing layer of the IoT architecture, different IoT applications in which PPTs are deployed, and the different privacy types addressed by these techniques within IoT systems. Key findings, such as the prominent privacy goal and privacy type in IoT, are discussed in this survey, along with identified research gaps that could inform future endeavors in privacy research and benefit the privacy research community and other stakeholders in IoT systems.
\end{abstract}


\end{frontmatter}

\section{Introduction}
\label{section:introduction}
\begin{figure}
        \centering
        \includegraphics[width=8cm,height=6cm]{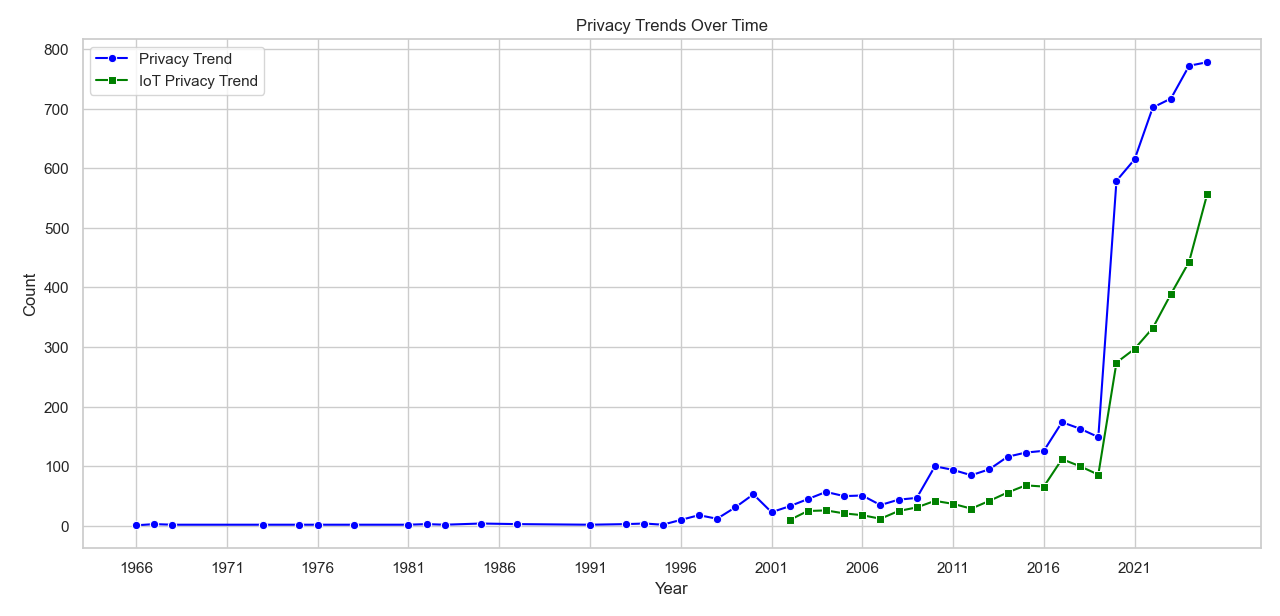}
        \caption{The line graph shows publication trends for PPTs. The blue line indicates the publications on general privacy, while the green line shows privacy-related publications in IoT systems.}
        \label{fig:line-chart}
    \end{figure}

This increase in privacy awareness is evident in the growing number of publications related to general privacy and those specifically focused on IoT systems from 1960 to early 2025, as shown in Figure \ref{fig:line-chart}. Figure \ref{fig:line-chart} was generated from an analysis of privacy-related publications collected from Google Scholar using Harzing's Publish or Perish software, designed to retrieve and analyze academic citations \cite{harzing}.

According to the analysis in this study, trends in IoT-related privacy publications began in 2002, as shown in Figure \ref{fig:line-chart}, with a significant increase from 2008 to 2025. This increase highlights increased awareness and recognition of the privacy risks associated with IoT systems.


As mentioned in some literature, technological advancements contribute to the new privacy challenges by exacerbating privacy risks \cite{gupta2023chatgpt}. Threat actors can now launch a privacy threat leveraging these new technologies, such as Generative AI \cite{gupta2023chatgpt, chen2024generative}. However, technological advancements have also facilitated the development of sophisticated PPTs \cite{cook2023security, meden2021privacy}. Some of these technologies, such as artificial intelligence (AI) and post-quantum cryptography, have been leveraged to build privacy-enhancing technologies (PETs) that are used to create PPTs, which mitigate privacy threats. A proper understanding of the PPTs development, application approach, and where it has been applied in IoT is essential to enlighten researchers on areas that require further research.

This insight motivates our study by leveraging a scoping review, a systematic method for mapping existing evidence on a topic, which identifies key concepts, theories, and knowledge gaps \cite{tricco2018prisma}. This study adopts this approach due to its comprehensive overview of a subject area, enabling researchers and practitioners to understand the current state of knowledge and identify areas for further investigation.

 Our aim is to use a scoping review to analyze various PPTs in the context of IoT, focusing on the technologies used to build PET, PPT implementation in the computing layers of IoT architecture, privacy protection types, different IoT applications in which PPTs are implemented, and privacy goals of these PPTs. This study addresses gaps not covered by previous reviews on PPT in IoT systems. The contributions of this study is summarized as follows:
\begin{itemize}
   \item A scoping review of state-of-the-art PPTs in IoT systems from paper publications in the past fourteen (14) years. 
    \item Analyzing studies on PPTs in the literature based on IoT application area, PPT goals, technologies used to build PETs, and the privacy types protected by these PPTs as gathered from the literature. 
   \item Proposing a clear relationship among PPT, PET, privacy threat, privacy goals, and privacy types.
   \item Presenting the importance of device identity privacy in IoT.
   \item A comparison of each PPT from the literature to identify the most frequently used and least used PPTs across different IoT applications. 
     
\end{itemize}

This paper is organized as follows: Section~\ref{section:introduction} explains the trends in general privacy publications and those specific to privacy in IoT systems. Section~\ref{section:Definitions} presents a brief background on privacy and key privacy terms in IoT systems, Section~\ref{subsec:Comparison} shows the uniqueness of this study by comparing it to previous reviews on PPTs. Section~\ref{section:Methodology} explains the scoping review guideline and framework used in this study. Section~\ref{section:RQ} identifies the research questions in this study, while Section~\ref{section:relevant study} explains how we gathered relevant studies in this scoping review. Section~\ref{section:Paper selection} presents the databases and the year range of selection of papers reviewed in this study. Section~\ref{section:Data Chart} categorizes the collected papers into five key areas our review focused on: PPT goals, technologies used to build PET, IoT applications and their PPTs, IoT computing layer in which PPTs are implemented, and privacy types in which these PPTs operate. Section~\ref{section:Summary} summarizes our findings and provides answers to the research questions in our study. Section~\ref{section:future} discusses the future direction based on the research questions in our study. Lastly, Section~\ref{section:Conclusion} presents the study's conclusions and its limitations.

\section{Overview of privacy in IoT systems}
\label{section:Definitions}
This section explores the fundamental concepts underlying privacy preservation within the Internet of Things (IoT) context. We begin by defining the concept of IoT and its architecture and elucidating the essence of communication modes in IoT. To fully appreciate the significance of PPTs in IoT, it is essential to establish a comprehensive understanding of privacy within the context of these techniques in IoT. Lastly, we examine the interplay between privacy terms such as PPT, PET, privacy types, privacy goals, and privacy threats in IoT systems.

This section will offer insights into the evolving landscape of PPTs in IoT, shedding light on the ever-changing dynamics of privacy and its safeguarding mechanisms.

\subsection{Internet of Things (IoT)}

IoT consists of "things" that can interact with each other by communicating and exchanging data and information based on the senses about the environment that trigger actions in the form of services with little or no human intervention \cite{sundmaeker2010vision}.

These "things" are connected sensors and actuating devices that form a flexible network that effortlessly shares information across different platforms. This system operates within a unified framework, encouraging the development of a shared operational perspective and driving the creation of advanced applications \cite{gubbi2013internet}. This progress is the result of combining ubiquitous sensing, data analytics, and information representation, with cloud computing being the central and cohesive foundation for this integrated ecosystem \cite{gubbi2013internet}. IoT was first presented in 1999 by Kevin Ashton, and with technology development such as AI, it has evolved into a platform that influences our day-to-day activities. Its application has spanned various areas, including medicine, industry, agriculture, education, and energy, making operations in these areas more intelligent \cite{balaji2019iot}. IoT systems comprise connecting IoT devices and smart devices, and are implemented using various architectures that cater to the specific needs of IoT applications.
\subsection{IoT Devices and Smart Devices}
IoT devices are physical objects in an IoT system that are connected to each other and to the internet to collect, send, receive, and store data. An example of an IoT device is a smart TV, a smart watch, a smartphone, to name a few. However, smart devices are intelligent objects, also known as smart objects, that possess unique identities, sensing and storage capabilities, data availability, communication abilities, and decision-making capabilities \cite{lopez2012using}. Smart devices are a subset of IoT devices and do not require human interaction. An example is a weather probe that collects weather data, and an IoT security system that involves a surveillance camera that communicates with other devices and sends alerts and footage in case of intrusion, leveraging Artificial Intelligence (AI) for asset tracking and monitoring, as seen in Industrial IoT (IIoT) \cite{elgazzar2022revisiting}. Although the data from these smart devices is eventually collected by humans for analysis. The analyzed data can be categorized as either user or device data.

\subsection{User and Device data in IoT System}
Various data in the IoT system are processed and stored, and then exchanged between IoT devices during their interactions. These data can be categorized as either user data or device data. 
\begin{itemize}
    \item User data: Data related to IoT users, which may be sensitive (e.g., personal identifiable information [PII]) or non-sensitive (e.g., device usage data) \cite{subahi2019detecting}. However, non-sensitive user data can become sensitive if linked to other data \cite{madaan2018data}. For example, smart switch usage in correlation with specific time periods of usage can infer sensitive user behavior data.

    \item Device data: These are data related to IoT devices. Some user data stored in these devices is often associated with device data. However, device data is typically information that can be used to identify an IoT device within an IoT system \cite{chowdhury2020network,guo2020detecting}. Examples of these devices include IP addresses, firmware versions, model number etc \cite{andrews2025iot, lei2025fine}.  
\end{itemize}

\subsection{IoT architecture}
The deployment of IoT systems is highly flexible and depends on the specific requirements and types of IoT applications being utilized. For instance, when implementing IoT in a Smart Home (SH) environment, a middleware architecture or a gateway for internet connectivity might suffice. On the other hand, in an industrial IoT setup, a cloud-based architecture integrated with both fog and edge computing may be necessary \cite{jamil2024enabling}. Consequently, we outline the four primary IoT architectures, as highlighted in existing literature, applicable across diverse IoT applications. 

\subsubsection{Locally Hosted IoT system}
In a traditional IoT system without cloud or Fog connectivity, a centralized software known as a \textbf{middleware} is essential to link the various IoT entities within the IoT ecosystem \cite{tawalbeh2020iot, zhang2021middleware}. This system offers a level of security when operating in a locally hosted environment, until the \textbf{Gateway}, an example of a middleware, permits external access to the Internet, making the locally hosted IoT system vulnerable to external threats \cite{zavalyshyn2022sok, demir2022secure}.

{\textbf{Middleware IoT system}}: It is an intermediary between an IoT system's devices and application layers \cite{zhang2021middleware}. It provides communication and interaction management services among IoT devices, platforms, humans, and applications. Middleware is an essential part of an IoT ecosystem that enables the easy management of IoT components and provides scalability, security, and reliability for IoT applications by delivering timely updates to the latest, secure version \cite{alalade2020intrusion1}. 
            
\begin{figure}
        \centering
        \includegraphics[width=8cm,height=6cm]{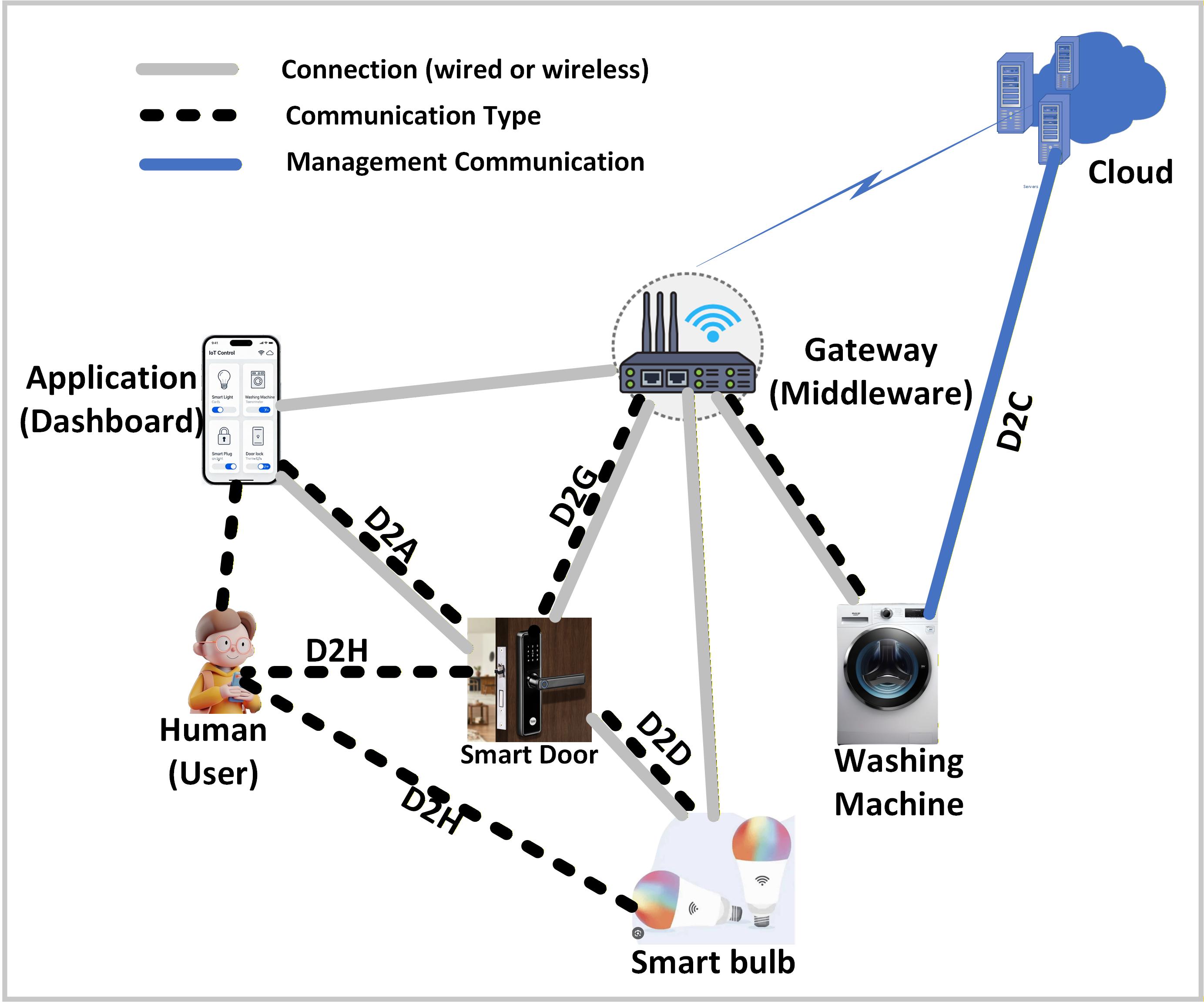}
        \caption{The architecture of a locally-hosted IoT system showing communication types between user, devices, gateway(middleware), and cloud}
        \label{fig:locally-hosted}
\end{figure}

 A typical middleware example is the \textbf{communication middleware} that allows communication management between IoT devices, platforms, and humans using different sets of protocols and standards for data exchange \cite{zhiliang2011soa}. Communication in IoT spans all entities that comprise the IoT ecosystem, as shown in \autoref{fig:locally-hosted}, and these types of communications are explained as follows.
     \begin{enumerate}[label=(\roman*)] \label{communication_types}
        \item \textbf{Device-to-device (D2D)}: This is the autonomous communication between IoT devices without any human intervention, with centralized control for gathering, sharing, and multi-hop forwarding of information \cite{bello2014intelligent, souri2022systematic}.
        \item \textbf{Device-to-cloud (D2C)}: This is the communication between the IoT device and the cloud service for storage purposes and management of the IoT system. The cloud can reliably store the IoT device data, with security, privacy, and uninterrupted availability \cite{alam2018cics, souri2022systematic, bello2013communication}.
        \item \textbf{Device-to-application (D2A)}: This is interaction between IoT applications and IoT devices. This application can be installed on a user's mobile device as an API manager or dashboard for controlling the IoT device or embedded into human life using content-centric, message technology, cloud technology, and sensor technology \cite{souri2022systematic}.
        \item \textbf{Device-to-gateway (D2G)}: This is usually a secure communication between the local gateway and the IoT devices \cite{souri2022systematic}. 
        \item \textbf{Device-to-Human (D2H)} : This the communication between human and the IoT device. This communication passes user information to the IoT device, and vice versa \cite{bello2014intelligent}.
    
     \end{enumerate}

       \subsubsection{Cloud-enabled IoT system} The IoT ecosystem recently depends greatly on cloud services through data storage, processing, and sharing support. This dependency is for adequate accommodation of ever-growing IoT data and service processing \cite{sharma2019cloud}. This allows data that cannot be stored locally on the IoT device due to limited storage capacity to be stored remotely in the cloud. The privacy of moving the data to the cloud and the privacy of the data at rest are essential \cite{zainuddin2021study}. Fog and edge computing are two common cloud-based approaches commonly used in IoT architecture, as shown in \autoref{fig:computing-layer}.

      \subsubsection{Edge-enable IoT system} Edge computing system serves as a bridge between the physical and digital realms as shown in \autoref{fig:computing-layer}. Devices at the edge gather real-world data, process it swiftly, and make it readily accessible to users via the internet \cite{Nasir}. In an edge-enabled IoT setup, the primary focus is on capturing sensor data and efficiently transmitting it to clients in a clear and understandable format \cite{Milenkovic2020}. Edge-enabled IoT offers scalability and enhances privacy in IoT systems by relocating applications from the cloud to edge servers, which are located closer to IoT devices.

      \subsubsection{Fog-enabled IoT system} Fog computing usually resides between cloud and edge computing, filtering out important data that needs to be processed in the cloud (see \autoref{fig:computing-layer}). A fog-enabled IoT system involves edge computing, which brings control, storage, and networking closer to the IoT devices \cite{pfandzelter2021zero}. This setup enhances user experience without replacing cloud computing; it works alongside it. Fog computing significantly improves the quality of service (QoS) compared to users directly accessing the cloud \cite{pfandzelter2021zero}. Privacy requirements in a fog-enabled IoT system, as explained by Kinza et al. \cite{sarwar2021survey}, revolve around content and context. Content privacy safeguards the data used for message delivery, while context privacy focuses on aspects of communication that could reveal the message's content. Communication within a fog-based IoT system involves conveying content and context, necessitating the protection of both device and user data \cite{sarwar2021survey}.

\begin{figure}
        \centering
        \includegraphics[width=8cm,height=6cm]{IoT-Architecture.jpg}
        \caption{An IoT architecture showing the computing layer of the IoT system}
        \label{fig:computing-layer}
\end{figure} 



\subsection{Privacy, Privacy Types, Privacy threats, Privacy Preservation Techniques(PPTs), General Technology and Privacy Enhancing Technologies (PETs)}
\label{privacy-taxonomy}

    \subsubsection{Privacy} It has come way back from the 14th century through the 18th, starting with one's body and home protection, until it evolved into personal information access control and protection at the end of the 19th century \cite{holvast2007history}. Privacy is essentially the ability of individuals to control how much and under what circumstances they reveal their thoughts and actions to others. This concept is closely linked to one's identity and is seen as a way to safeguard an individual's autonomy and independence \cite{kris}.  Privacy encompasses an individual's right to control the collection, storage, and disclosure of their personal information. It serves to protect data, control how long data is retained, and regulate its use \cite{alhirabi2021security, seliem2018towards}. When unauthorized exposure of this information occurs, it is known as a privacy breach. In the IoT context, a privacy breach refers to the disclosure of information about IoT users and devices to external entities, thereby posing a threat to individuals' privacy \cite{ogonji2020survey}.

    \subsubsection{Types of Privacy Protection}
   Privacy in the IoT revolves around users and other entities interacting within IoT systems. Finn et al. \cite{finn2013seven} identify seven types of privacy in technology. These types include the privacy of the person, the privacy of personal behavior, the privacy of personal communication, the privacy of personal data, the privacy of location and space, the privacy of thoughts and feelings, and the privacy of associations \cite{finn2013seven}. However, not all of these privacy types are relevant to IoT. Therefore, based on existing literature, this study outlined the specific types of privacy in IoT systems.
     
    \begin{itemize}
        \item \textbf{User privacy}: To protect user data by the service provider and provide transparency on the kind of users' information they are collecting \cite{liyanage2018comprehensive}.
        \item \textbf{Device Identity privacy}: To prevent unauthorized device tracking and smart device information disclosure, such as firmware version, operating system, MAC address, etc. \cite{liyanage2018comprehensive}.
        \item \textbf{Communication channels privacy}: To protect communication link conveying information between device-to-device, device-to-cloud, device-to-human and device-to-application \cite{finn2013seven}.
        \item \textbf{Location privacy}: To protect the user and device location information. This could prevent threats such as profiling \cite{finn2013seven, liyanage2018comprehensive}.
        \item \textbf{Group privacy}: To protect the privacy of a group of users \cite{finn2013seven}. For example, a family member of a Smart Home (SH) owner or department member in an Industrial IoT (IIoT) environment who has access to operate an automated system in an industry.
    
    \end{itemize}
     \subsubsection{Privacy Threats in IoT}
      Depending on the threat actor, some privacy threats can be malicious, while some can be non-malicious. Privacy threats, such as subject identification, linkage, data leakage, inventory attacks, eavesdropping, life cycle transition, privacy violations during presentation, and data tampering, can be malicious due to the threat actor involved \cite{ogonji2020survey, ziegeldorf2014privacy, zainuddin2021study, seliem2018towards}. In contrast, privacy threats such as Localization and tracking, and profiling are not usually malicious when they involve actors such as service providers, security agents, and third-party providers \cite{ogonji2020survey, ziegeldorf2014privacy}. These privacy threats could affect either IoT users or devices, or both

     
    \subsubsection{Privacy Preservation Techniques (PPT)}
    PPTs are methods or processes implemented in IoT to achieve privacy goals \cite{shimona2020survey}. These privacy goals vary depending on the techniques and the IoT application involved. These privacy goals include data masking,  pseudonymization, data perturbation, multi-path routing, and access controls. To achieve this goal by PPTs, the right technological building block needs to be developed, which is referred to as privacy-enhancing technologies (PET) \cite{sarwar2021survey, abi2018preserving, torre2023privacy, akil2020privacy}. PET in IoT may be challenging due to the constraints of IoT systems, such as limited memory capacity and processing power. However, implementing these PETs in a lightweight manner will make them suitable for IoT systems \cite{shimona2020survey}.
    
    \subsection{ Privacy-Enhancing Technologies (PET)}  

    These are technologies used to achieve PPTs in an IoT system. In IoT, PPTs have evolved, employing diverse PETs (such as differential privacy, homomorphic encryption, and anonymization, to mention a few) to provide various privacy preservation solutions \cite{li2022flexible, sezer2023ppfchain, chamikara2021privacy, hindistan2023hybrid, li2022privacy, salim2022perturbation, kong2019privacy}. 

    \subsection{General Technology used in privacy Privacy Preservation Approach in IoT}  \label{sec:general technology}
    The Internet of Things (IoT) utilizes various privacy preservation approaches to achieve specific privacy goals and types. These approaches often leverage PETs developed through general technological approaches like blockchain \cite{ding2023enhancing}, machine learning/deep learning \cite{zhu2023enhancing}, cryptography \cite{ali2023healthlock}, informed consent \cite{gheisari2023agile}, quantum-safe algorithms \cite{samonte2024integrating}, mathematical models \cite{palekar2024apro}, and privacy-based infrastructure.

\begin{figure*}[ht]
  
        \centering
        \includegraphics[width=13cm,height=7.5cm]{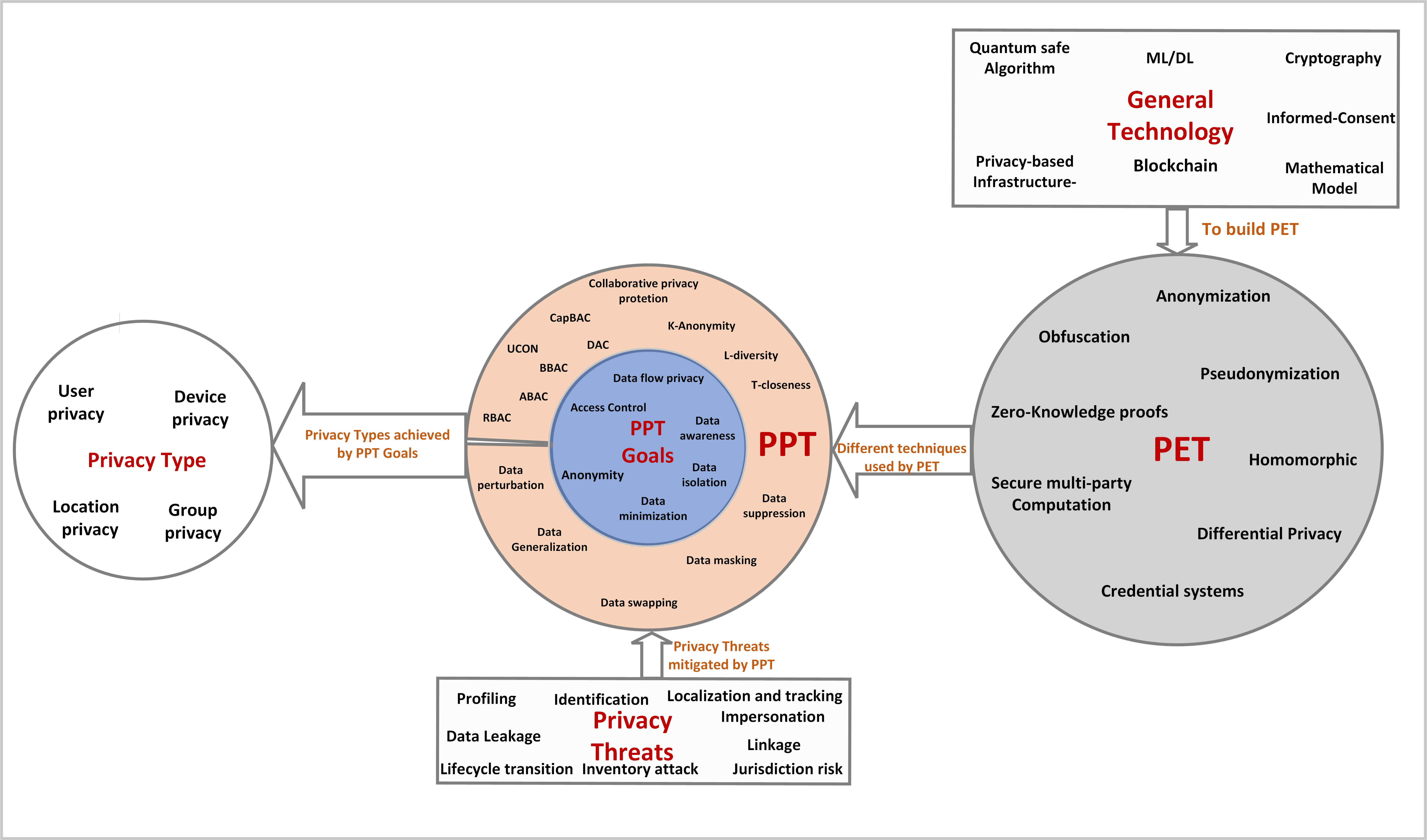}
        \caption{Privacy terms and their relation based on privacy preservation in IoT}
        \label{fig:PP-in-IoT}

        \footnotesize  
        \textit{\textbf{Note:}  Existing and emerging technologies are utilized to build PET, which is essential for developing PPTs that achieve specific privacy goals in IoT. These privacy goals provide different privacy types. The overall process provides privacy preservation that mitigates privacy threats in IoT systems.} 
        
\end{figure*}

 The overall privacy preservation taxonomy and the involved entities are represented in \autoref{fig:PP-in-IoT}   
    
  


\section{Comparison of this paper with other surveys and review papers}
\label{subsec:Comparison}

In this section, we comprehensively analyzed areas previously explored by surveys focusing on PPTs within IoT. These areas were classified based on the technology implemented, the goals of PPTs, the specific IoT applications in which these techniques were applied, and the types of privacy they aimed to protect. The comparative insights shown in Table \ref{table:study-comparison} highlight the uniqueness of this scoping review. This study is the first to conduct a comprehensive scoping review of PPT in an IoT system, providing a detailed understanding of the technologies used to build PETs, IoT applications, privacy goals, the implementation of PPT across different layers of IoT computing, and the privacy types associated with these techniques. We present an explanation of similar reviews related to this study as follows:

\begin{table*}[ht]
    \captionsetup{justification=centering}
    \centering
    
    \normalsize 
    \setlength{\extrarowheight}{2pt}
    \captionof{table}{Comparison of Our work to Prior Reviews on  Privacy Preservation in IoT}
\label{table:study-comparison}
\fontsize{9pt}{9pt}\selectfont
    \begin{tabularx}{\textwidth}{|>{\raggedright\arraybackslash}m{1.0cm}
                                 |>{\centering\arraybackslash}m{1.5cm} 
                                 |>{\centering\arraybackslash}m{1.0cm}
                                 |>{\centering\arraybackslash}m{2cm}
                                 |>{\centering\arraybackslash}m{1.1cm}
                                 |>{\centering\arraybackslash}m{3.5cm}
                                 |>{\centering\arraybackslash}X|}
  
 \hline
 Paper & PPT Review & PPT Goal & IoT application & Privacy Type & PPT Implementation in IoT Computing Layers &  Literature Review\\
 \hline
 \cite{sarwar2021survey}   & \textcircled{2} &  \halfcirc & \fullcirc & \fullcirc &  \halfcirc & Systematic Mapping \\
\hline
\cite{abi2018preserving}  &  \textcircled{2}  &  \halfcirc & \fullcirc &   \halfcirc &   \halfcirc & Narrative Review\\
\hline
\cite{torre2023privacy}  & \textcircled{2} &  \halfcirc & \fullcirc &  \emptycirc &  \emptycirc & Systematic Mapping\\
\hline
\cite{akil2020privacy} & \textcircled{2} &  \halfcirc & \fullcirc &  \emptycirc &  \emptycirc & Systematic Review\\
\hline

Our work & \fullcirc \textcircled{1} \textcircled{3} & \fullcirc & \fullcirc \textcircled{3} & \fullcirc & 
 \fullcirc & Scoping Review\\
\hline

\end{tabularx} 

 \footnotesize       
        \textcircled{1}: The study covers both technical and non-technical PPTs approach $~~$   \\
        \textcircled{2}: The study covers only the technical privacy preservation technique approach. \\
        \textcircled{3}: The study introduces new topics and emerging technologies \\
        \fullcirc: The study reviews the topic comprehensively,  $~~$
        \halfcirc: The study covers the topic partially, \\
        \emptycirc: The study does not cover the topic.  \\
        
\end{table*}

Kinza et al. \cite{sarwar2021survey} conducted a study on how privacy can be protected in IoT applications. They placed a particular emphasis on how fog computing can strengthen privacy in the IoT field. They discussed the technology used for implementing privacy solutions which are anonymization and data access control. Their study primarily focused on technical approaches, such as homomorphic encryption, logistical scrambling, and distributive role-based access controls \cite{sarwar2021survey}. Additionally, the researchers clearly explained the types of privacy protection covered by the techniques they examined. While they reviewed many IoT applications, the specific privacy objectives of the studied techniques received relatively limited attention.

In their review, Abi Sen et al. \cite{abi2018preserving} partially discuss some technologies of use in their study, primarily focusing on mathematical models and cryptography used for privacy preservation in IoT systems in the literature. However, their study covers common privacy goals and IoT applications (except newly introduced IoT applications in multimedia, social networks, and wearables). They primarily emphasized the preservation of user and location privacy as the central focus of their review, highlighting the challenges associated with these specific privacy protection types in their study.\cite{abi2018preserving}. Additionally, they explain the significance of central cloud storage in IoT and its inherent limitations, proposing the adoption of Fog computing to address these limitations through decentralized storage and processing capabilities.

Torre et al. \cite{torre2023privacy} address a few specific privacy goals of privacy techniques reviewed in their study. These goals include anonymization, access control, data flow privacy, data minimization, and obfuscation. While the authors \cite{torre2023privacy} provide some insights into the technology used for privacy preservation, their primary focus centers on cryptography. However, they also touch upon blockchain and machine learning as pertinent approaches. Furthermore, their review predominantly focuses on IoT applications in health, transport, and wearable \cite{torre2023privacy}.

In their systematic review, Ali et al. \cite{akil2020privacy} focus on privacy-preserving identifiers in IoT environments, particularly those that enable pseudonymity. The primary focus of the study lies on X.509 certificates for cryptography and privacy identifiers that facilitate pseudonym usage and data minimization. Their study of IoT applications is domain-specific, with a specific focus on areas such as health, smart homes, transportation, and energy production \cite{akil2020privacy}.

Some gaps were observed in previous reviews on PPTs in the IoT environment, especially in the technology used to build PETs, privacy goals, privacy preservation types, the IoT application of interest, and the PPT implementation in the IoT computing layer. Prior reviews on PPT in IoT focus on studies that use technical approaches to achieve PPT. However, PPTs in IoT should also incorporate non-technical approaches such as informed consent. Furthermore, this study discovered that no prior reviews have covered new trends in IoT applications in areas such as social networks and multimedia.

We addressed some of the research gaps by discussing potential solutions (see \autoref{section:future}) in response to specific research questions in \autoref{section:RQ}. Our study also examined the objectives of various privacy techniques, including hybrid solutions that integrate multiple approaches. Additionally, we focused on various types of privacy, particularly those that have received less attention from researchers, such as device identity privacy, and analyzed IoT application areas, specifically those recently introduced in the context of IoT.

\section{The Five-step Research Methodology}
\label{section:Methodology}
We adopted PRISMA-ScR (Preferred Reporting Items for Systematic Reviews and Meta-Analyses Extension for Scoping Reviews) due to its comprehensive search strategy, which helps map the literature into meaningful categories within a broad context. The scoping review reporting guideline was initially introduced by Tricco et al. \cite{tricco2018prisma}. This guideline features a detailed checklist and explanations of PRISMA-ScR. This study implements PRISMA-ScR within the framework established by Arksey et al. \cite{arksey2005scoping}. Arksey's framework consists of five stages: (1) identifying research questions, (2) identifying relevant studies, (3) selecting pertinent studies, (4) charting the collected data (data visualization), and (5) synthesizing and presenting the findings. It is essential to note that this guideline inherently encompasses the key components outlined in the PRISMA reporting methodology.
\section{Step 1: Identifying the research questions}
\label{section:RQ}
To better understand the factors that affect privacy and its preservation in IoT systems, we formulated essential questions that have not been addressed in previous studies.

These inquiries reveal what, why, and how certain IoT activities and applications impact the techniques used to preserve privacy in IoT systems. Here, the study presents a set of research questions (RQs) that are not adequately addressed in prior reviews on PPTs in the IoT. These research questions are (1) What technologies are prominently used to develop PETs in IoT systems? (2) What are the dominant PPT's privacy goals in IoT applications? \label{RQ2}, (3) What are the commonly known and recent IoT applications added to the literature?, (4) What computing layer in IoT architectures are PPTs mostly implemented within IoT systems, and which IoT application has PPT most implemented from the literature?,(5) What privacy types are the main focus of most PPTs in IoT systems?

\section{Step 2: Identifying relevant studies}
\label{section:relevant study}
To understand privacy preservation in the IoT environment, we queried selected databases with the following search expression: [("Privacy preservation techniques" OR "Privacy protection techniques" OR "Privacy preservation methods" OR "Privacy protection Approaches" OR "Privacy protection methods" OR "Privacy protection techniques") AND ("Internet of Things" OR "IoT" OR "Smart Things" OR "Smart environment" OR "Internet of Everything")].

\begin{figure}
        \centering
        \includegraphics[width=8cm,height=10cm]{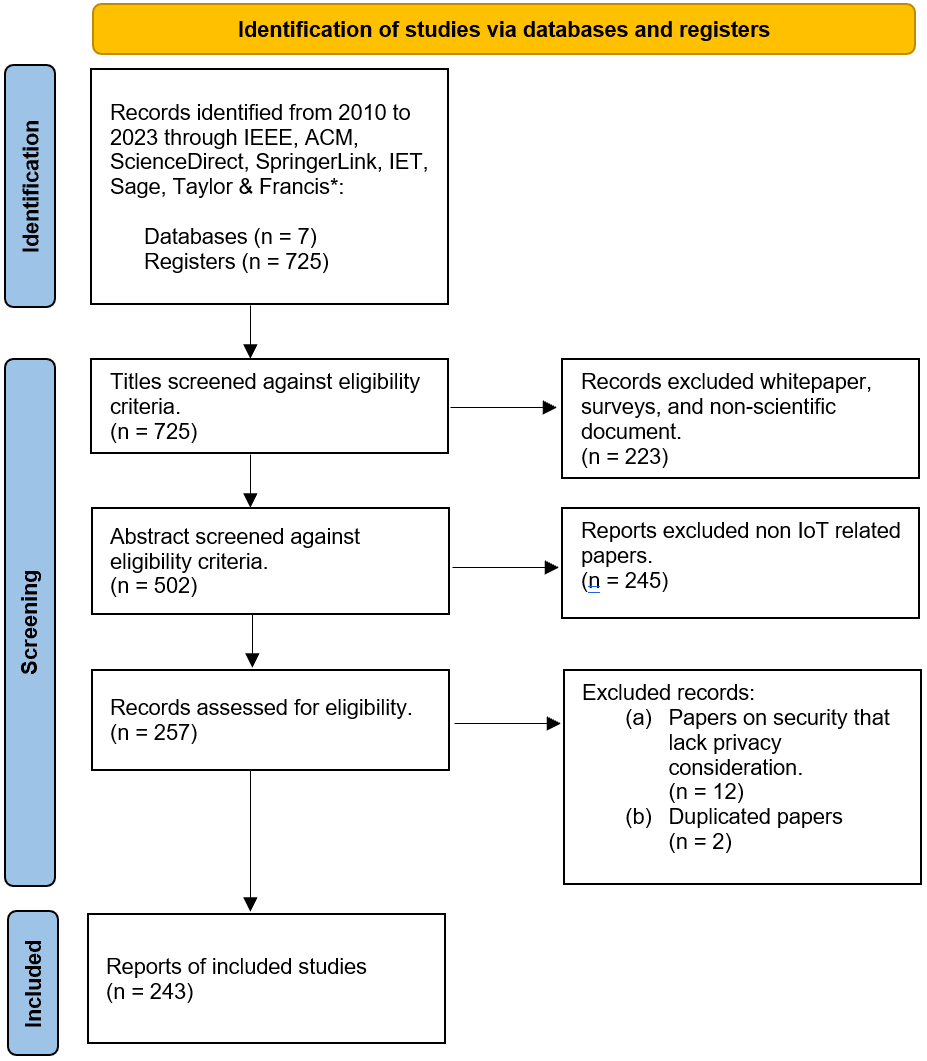}
        \caption{The flow diagram showing the paper selection criteria}
         \label{fig:PRISM}
    \end{figure}

Our work relied on reputable sources from seven databases: IEEE Xplore, ACM Digital Library, ScienceDirect, SpringerLink, Sage Journal, Wiley Online, and Taylor \& Francis, ensuring the robustness and reliability of our findings. As illustrated in Figure \ref{fig:PRISM}, a total of 1177 papers were initially collected, including conference proceedings, journal articles, white papers, websites, and theses. The first step of this study involved reviewing the titles, during which 502 papers were excluded for reasons such as being white papers, surveys, non-academic documents, or articles with irrelevant content. Next, the abstracts and conclusions of the remaining 675 papers were examined, excluding 327 papers that were not related to the Internet of Things (IoT). From the remaining 348 papers, twelve (15) papers that lacked privacy considerations were excluded, and four duplicates were removed. This process resulted in a final total of 329 papers for our review.

We applied both inclusion and exclusion criteria. The inclusion criteria required papers to focus on PPT in IoT, propose at least one PPT for IoT, be peer-reviewed, and be published between 2010 and 2025. However, the exclusion criteria ruled out secondary manuscripts, non-IoT-related papers, non-educational papers, white papers, non-scientific documents, and non-English papers.

 \section{Step 3: Selection of PPT publications in IoT}
 \label{section:Paper selection}
 Seven databases were investigated to gather publications relevant to PPTs in IoT systems. IEEE Xplore was the most prolific, contributing about 46\% of the total publications. In contrast, Taylor \& Francis, Sage, and Wiley Online had the lowest contributions, as shown in Figures \ref{fig:Journal-Database-bar} and \ref{fig:Journal-Database-pie}.

The volume of PPT publications collected increased from 2010 to 2018, reached its peak in 2019, but declined in 2020 and 2022, with another notable peak in 2021. In 2023 and 2024, there were extensive collections of papers; however, early December of 2025 marked the end of our paper collection, which had fewer publications than in 2024. We expect more publications to be released before the end of 2025, as illustrated in Figure \ref{fig:year-of-publications}.

Figure \ref{fig:PPT-Journal-Database} shows that IEEE Xplore houses the highest number of PPT papers based on PPT goals in IoT systems, followed by ScienceDirect. Taylor \& Francis, Wiley Online, and Sage contributed modestly, each offering between one and two PPT papers in the field of IoT.

 \subsection{Selection of PPTs publication in IoT}
\begin{figure}[ht]
        \centering
        \includegraphics[width=7cm,height=5.5cm]{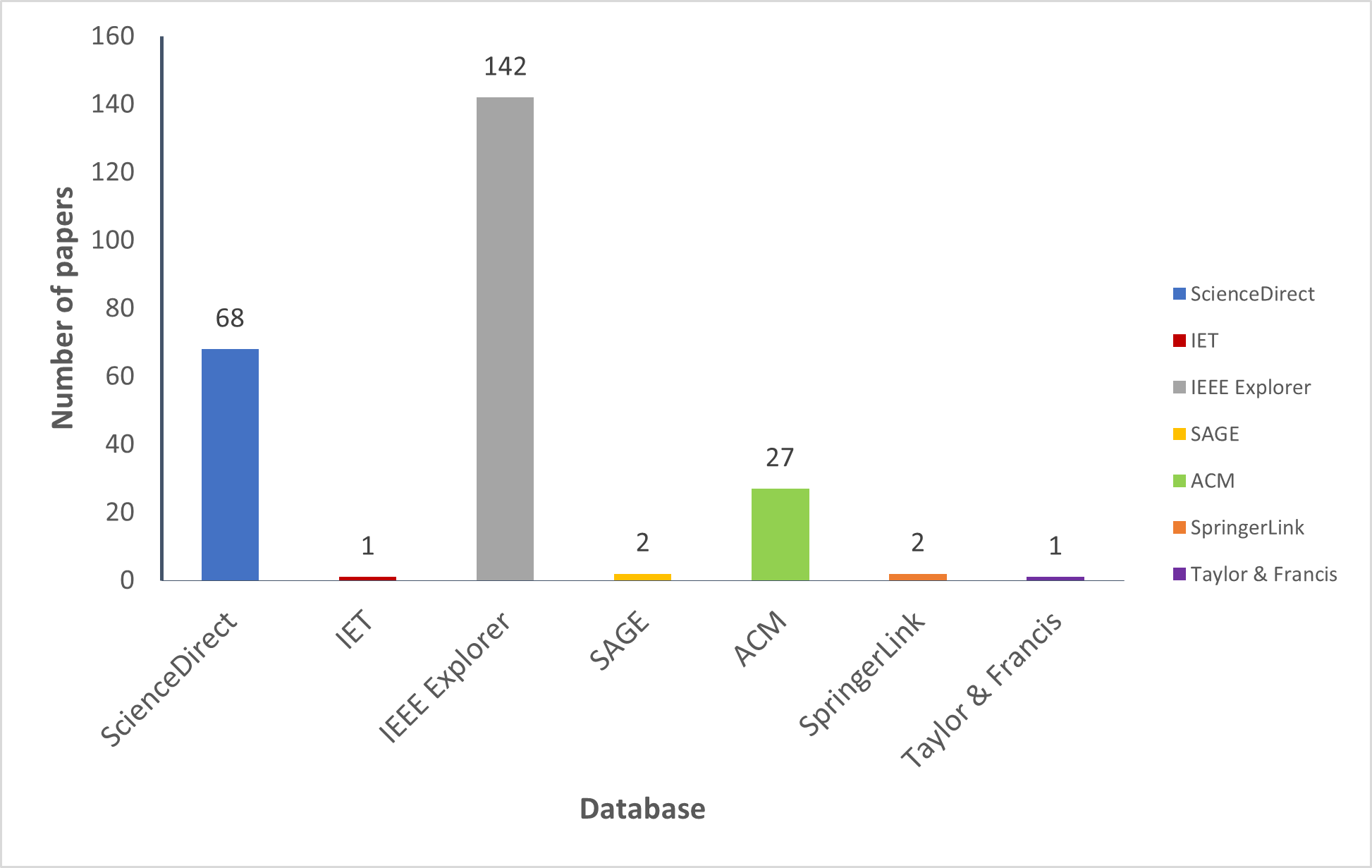}
        \caption{Number of papers in Database on PPTs in IoT Applications considered in our review}
        \label{fig:Journal-Database-bar}
\end{figure}

\begin{figure}[ht]
        \centering
        \includegraphics[width=7.5cm,height=5cm]{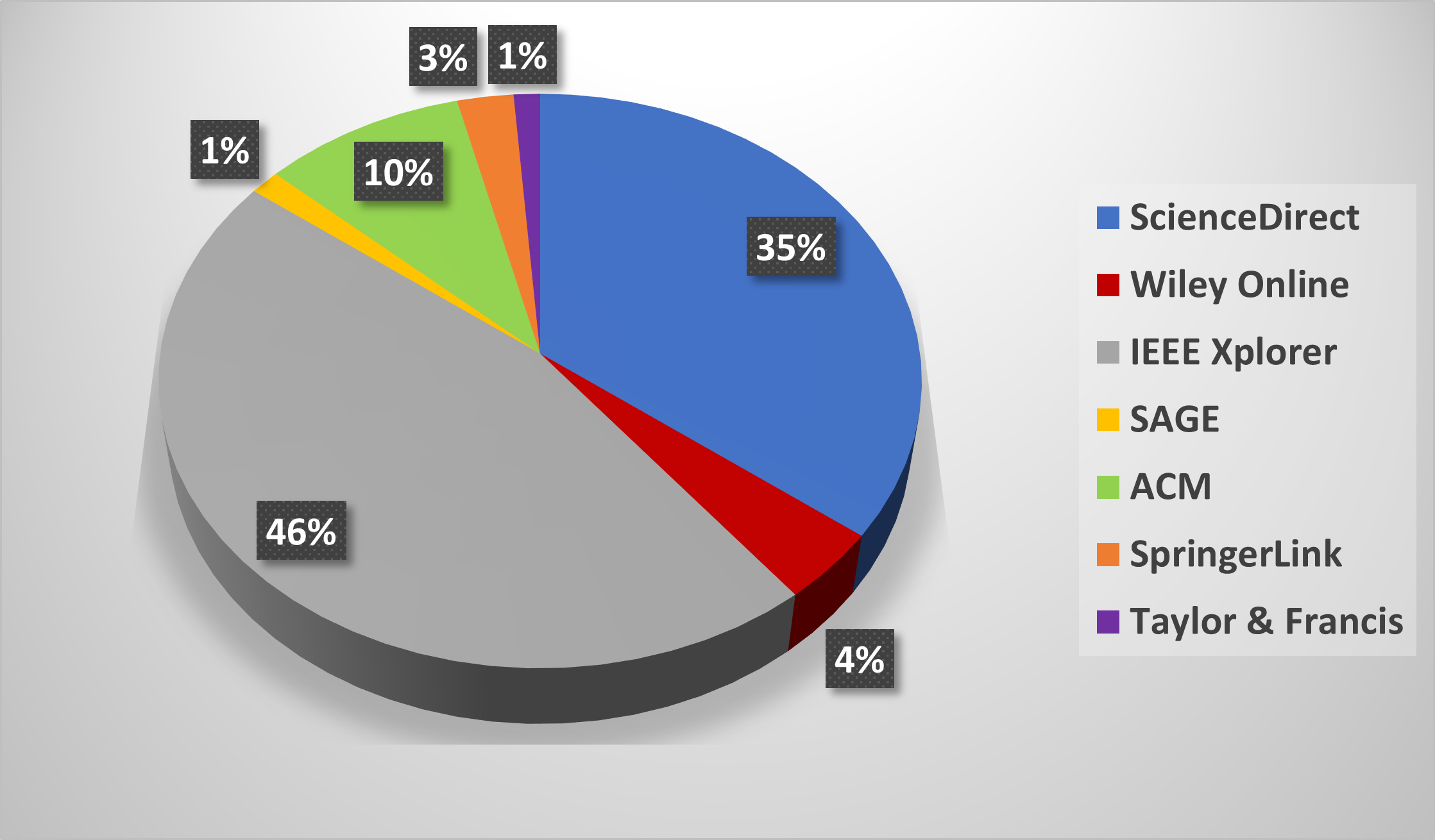}
        \caption{Percentage of papers in Database on PPTs in IoT Applications considered in our review}
        \label{fig:Journal-Database-pie}
\end{figure}

Our study examined seven(7) databases to explore their content on PPTs in the IoT system. Among these databases, IEEE Explorer emerged as the most prolific, boasting approximately 59\% of the total publications on the subject. However, Taylor and Francis, Sage, and IET exhibited the lowest contribution, each accounting for a mere 0\% of the publications encompassed in our review, as illustrated in both Figure \ref{fig:Journal-Database-bar} and \ref{fig:Journal-Database-pie}.

\begin{figure}[ht]
        \centering
        \includegraphics[width=8cm,height=6cm]{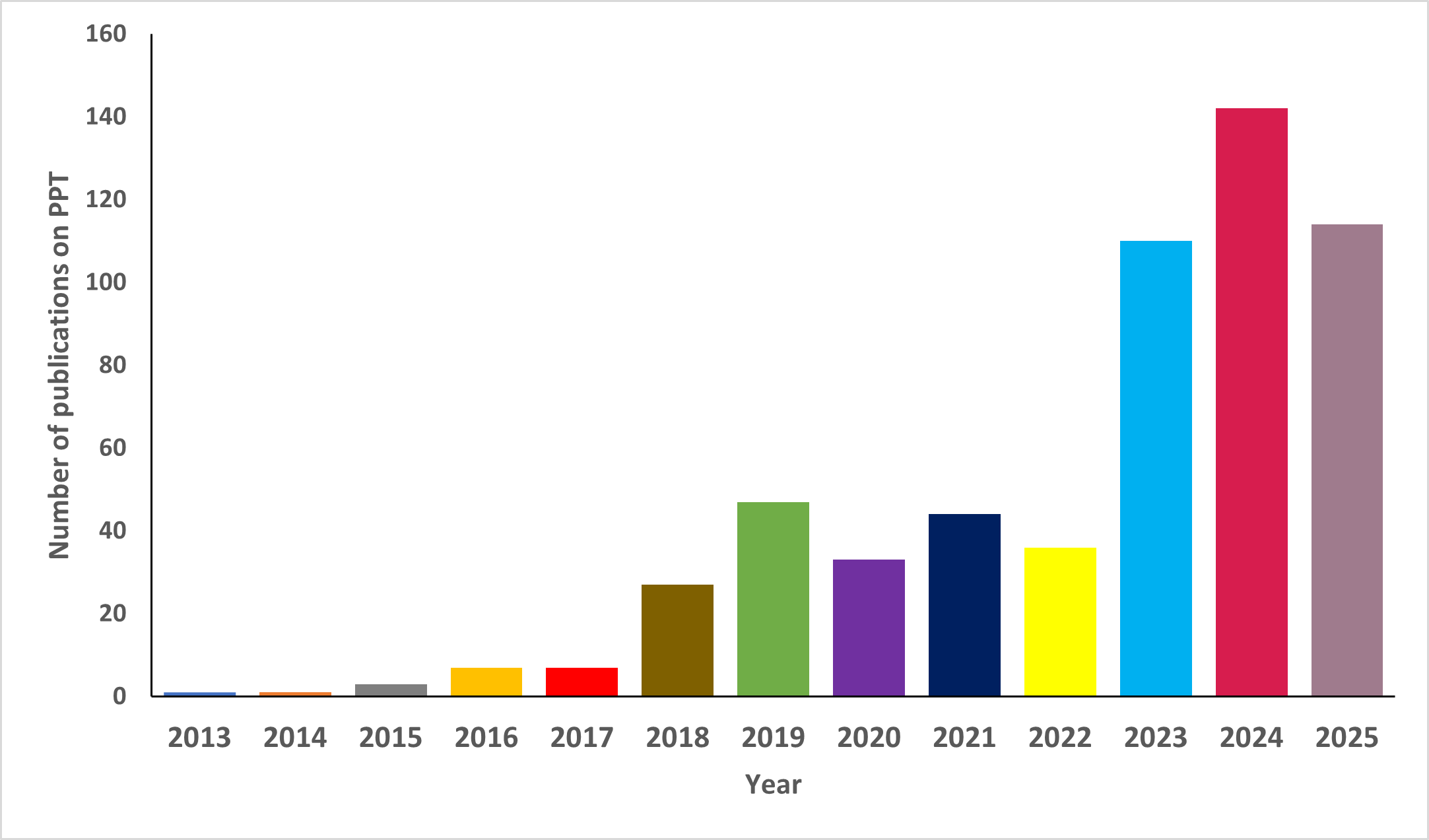}
        \caption{Number of publications on PPTs in IoT per year between 2010-2025}
         \label{fig:year-of-publications}
\end{figure}

To adhere to our exclusion criteria, we confined our focus to journal reviews published between 2010 and 2023. Notably, the early years from 2013 to 2017 witnessed a dip in publications, with 2019 marking the zenith in publication volume. Subsequently, from 2019 to 2023, it experienced a publication surge, with another notable peak in 2021. While collecting journals in April 2023, we amassed 31 journals published in 2023, indicating a noteworthy increase in journal publications on PPTs in the first quarter of that year, as highlighted in Figure \ref{fig:year-of-publications}.

\begin{figure*}[ht]
        \centering
        \includegraphics[width=14cm,height=8cm]{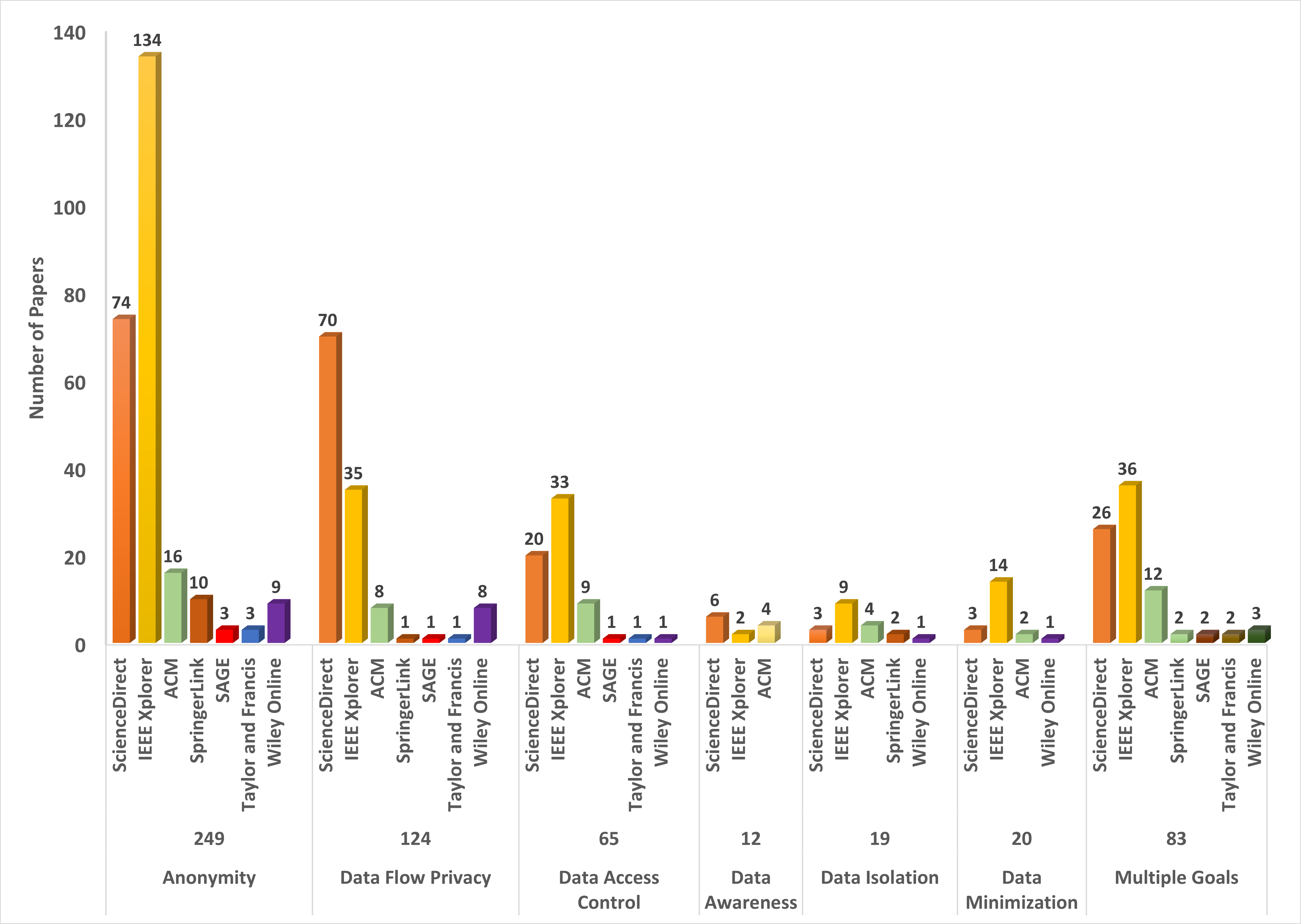}
        \caption{Papers on PPT goals in IoT systems and their published Database}
       \label{fig:PPT-Journal-Database}
\end{figure*}

Furthermore, Figure \ref{fig:PPT-Journal-Database} elucidates that the IEEE Explorer database houses the highest number of journals on PPTs categorized based on PPT goals in IoT systems, with ScienceDirect following closely as the second-highest database featuring an extensive collection of journals on PPTs in IoT. In contrast, Taylor and Francis, The Institute of Engineering and Technology (IET), and Sage contributed modestly, each offering between one and two journals on PPTs in IoT.
 
  \section{Step 4: Charting the data collected on PPT publications}
  \label{section:Data Chart}
The collected papers are categorized by PPT goals, technology used for building PET, IoT applications utilizing PPTs, computing layers where PPT is integrated within the IoT architecture, and the type of privacy protection. Our results address our research questions in subsection \ref{section:Summary} by reviewing each category as follows.

\subsection{PPT and PPT Goals}
\label{subsection:PPT Goals}
Many studies on IoT privacy have introduced various techniques aimed at different privacy goals. To organize this extensive literature, publications were categorized from various databases by the specific privacy protection goals they addressed.

\subsubsection{Anonymity} Anonymity in data privacy involves altering the original data format to hide sensitive parts of the data that could lead to the identification of the data subject \cite{george2013data, murthy2019comparative, salas2018some}. Various techniques are used to achieve anonymity:
\begin{itemize}
        \item Data perturbation: Adding random noise to data to preserve its usefulness for analysis while protecting its sensitive content \cite{geetha2012non}.
        
         \item Data generalization: Reducing data precision to a level that prevents easy linkage to individuals. It is used to maintain data accuracy for tasks such as medical data analysis \cite{wang2004bottom, Dalvi2015BottomUpGA}.
        
         \item Data masking: Concealing original data through modified content, retaining the data format and utility while protecting sensitive information \cite{tachepun2020data,aslanyan2019privacy}.
         
         \item Data swapping: Exchanging data attributes' values with values from different records to preserve statistical characteristics while obscuring individual data records\cite{hasan2016effective} \cite{mivule2017data}.

         \item  Pseudonymization: Replacing identifiable information with pseudonyms or codes to protect sensitive data, making it difficult to link transformed data back to original identities \cite{rottondi2012data} \cite{neubauer2009evaluation}.
    
\end{itemize}

    \subsubsection {Data Flow Privacy} This protects sensitive data during transmission processes such as data offloading, sharing, routing, and collaboration.

    \begin{itemize}
         \item Data sharing protection: Data sharing is essential in IoT systems, but it also poses significant privacy risks. Different PPTs are employed to protect sensitive data during transmission and sharing processes \cite{bhatia2016privacy}.
         
         \item Multipath routing:  IoT networks like mobile ad hoc networks (MANETs) use multipath routing protocols to enhance privacy protection for location, identity, data, and traffic as proposed by Taheri et al. \cite{taheri2013anonymity} 
    
         \item Data verification protection: This process reviews and helps improve existing privacy implementation in an IoT system. Techniques such as Petri net analysis, proposed by Malek et al. \cite{ab2016privacy}, are used to analyze and improve privacy implementations in IoT systems, identifying and addressing privacy violations.
    
         \item Route path protection: This is crucial in location-based services in IoT networks.
         For example, Yang et al. \cite{yang2012path} propose obfuscating the path to protect users' path privacy.
         \item Collaborative privacy protection: Collaboration in a distributed architecture is essential. However, trust needs to be established among the entities involved. Therefore, privacy techniques that collectively protect data sharing during collaboration are crucial for establishing trust among collaborative entities in the IoT environment \cite{zhang2021user}.

     \end{itemize}
     
    \subsubsection{Data Access Control} Access control in IoT systems involves allocating and granting access privileges based on defined policies and models. Ragothaman et al. \cite{ragothaman2023access} discuss eight access control models, with seven being the main focus in PPT papers as they are associated with IoT systems:
    
       \begin{itemize}
         \item Role-Based Access Control (RBAC): Users are granted access based on their roles within the system, ensuring appropriate access privileges. Managing RBAC in large and dynamic IoT environments can be challenging \cite{ragothaman2023access}\cite{ouaddah2017access}.
         
         \item Attribute-Based Access Control (ABAC): The access privilege is based on attributes or features of an IoT device, such as device ID, location, and IP address, to mention a few, and can be used for evaluating the device before access is granted \cite{ragothaman2023access} \cite{ouaddah2017access}.
         \item Blockchain-Based Access Control (BBAC): 
          Access control based on blockchain technology includes transaction-based and smart contract-based mechanisms. Transaction-based control enables the granting, revoking, or delegating of access, while smart contracts enforce access policies \cite{ragothaman2023access}.

         \item Capability-Based Access Control (CapBAC):  This approach requires the user or device to possess certain capabilities before the access token is delivered. This approach does not rely on the context before providing access \cite{ouaddah2017access}.
         \item Usage-Based Access Control (UCON):  UCON integrates authorization, obligation, and condition concepts. Authorization checks user eligibility, obligation specifies conditions for access, and conditions define criteria that users must fulfill \cite{ragothaman2023access} \cite{ouaddah2017access}.
 
         \item Relationship-based Access Control (ReBAC): ReBAC grants access based on established relationships in IoT systems, such as user-to-user or device-to-device relationships. This approach supports dynamic Identity Relationship Management (IRM) solutions \cite{ragothaman2023access}.

         \item Discretionary Access Control (DAC): DAC grants access based on an Access-Control List (ACL) managed by administrators. It requires careful monitoring and updating, especially when adding or removing devices or users from the IoT network \cite{ragothaman2023access}\cite{ouaddah2017access}.

     \end{itemize}
     \subsubsection{Data Awareness} Ensuring data subjects are informed and have control over their data, aligning with General Data Protection Regulation (GDPR) policies on user-informed consent \cite{GDPR}. For example, Tamani et al. \cite{Tamani} proposed using a semantic firewall infrastructure to filter personal data access in IoT.

     \subsubsection{Data Isolation} This process involves confining IoT devices or services within their own network segment. Thapa et al. \cite{thapa2018mitigating} describe different levels of isolation, and Zhang et al. \cite{Zhangvirtual} proposed virtualization and containerization in smart healthcare systems.

     \subsubsection {Data Minimization} Reduces data collection and processing in IoT systems to ensure only relevant data is used, preventing over-utilization. Studies focus on data minimization in IoT include \cite{abdulzahra2020data, singh2020embedded, osia2020hybrid}.

     \subsubsection{Multiple Goals} Some PPTs aim to achieve multiple goals simultaneously: Data anonymization and access control in \cite{qashlan2021privacy, yang2023anonymous, agrawal2022security}, data flow privacy and access control in \cite{peng2021efficient}, data flow privacy and minimization in \cite{deebak2022ai}, data isolation and distortion in \cite{salim2022perturbation}. Figure \ref{tab: multiple goals} presents other papers that achieved multiple goals in their proposed work.

    \textbf{Our observations}: From our study, anonymity emerges as the primary goal of PPTs across IoT application systems. Approximately 52\% of publications employ anonymity, with data perturbation the main privacy goal, followed by data masking and data pseudonymization. This observation is supported by our findings, as shown in Table \ref{tab: PPTGoal table}. Additionally, many PPTs address multiple privacy goals simultaneously, as shown in Table \ref{tab: multiple goals}. We observed an increase in the use of the PPT approach targeting multiple goals. The number of publications that achieve both anonymity and data flow protection increased from 2023 to 2025, as shown in \autoref{tab: multiple goals}. These findings underscore the importance of achieving anonymity in protecting sensitive data across diverse IoT applications while highlighting the need for comprehensive privacy strategies that simultaneously address multiple facets of data protection.

\begin{table*}[htbp]
\caption{Papers on PPT Goals and their Approaches}
\label{tab: PPTGoal table}
\fontsize{8pt}{9pt}\selectfont
\centering

\begin{tblr}{
  width = \linewidth,
  colspec = {Q[l,wd=0.12\linewidth] Q[l,wd=0.12\linewidth] X[j] Q[c,wd=0.15\linewidth]},
  cell{2}{1} = {r=5}{m},
  cell{7}{1} = {r=5}{m},
  cell{12}{1} = {r=7}{m},
  vlines,
  hline{1-2,7,12,20-23} = {-}{},
  hline{3-6,8-11,13-19} = {2-4}{},
}
\textbf{PPT Goals} & \textbf{PPT approach} & \textbf{Paper} & \textbf{Number of Paper} \\

Anonymity & Data masking & \cite{guo2020new, xu2020aggregate, gope2018lightweight, halder2023radio, nasiraee2020anonymous, xia2021effective, chen2023bcgs, satyanarayana2023comparative, deng2022secure, wang2018privacy, elmisery2016fog, yang2018density, wang2018blockchain, zhang2019new, zhu2019secure, chen2019light, rodriguez2020cooperative, ming2020efficient, ahmed2020anonymous, shin2020privacy, wan2020internet, yu2021lightweight, alzahrani2021provable, liu2023situ, li2022data, chaudhry2021rotating, wu2019lightweight, tan2019secure, zhang2019provable, xie2022improved, zhang2018tolerating, chintan, wang2021forward, wang2022panda, fang2022privacy, yin2022novel, meng2019steganography, sundarakantham2023hybrid, liu2023improvement, elmisery2017cognitive, al2022counterfeit, yan2021scalable, yan2017context, baccour2021rl, pistono2021cryptosystem, xu2022cryptoanalysis, qiu2019efficient, waheed2023privacy, yang2023privacy, gheisari2023agile, chang2023practical, karthikeyan2024privacy, jibril2024semantic, li2025aeppfl, 2025rehan, dong2025dummy, xu2025privacy, yenugula2025privacy, marchioro2023practical, hua2024international, bigelli2024privacy, deepak2024privacy, shen2025privacy, mahendran2025prism, wang2025privacy, zhao2025reversible, jaganraja2025agile, odeh2025secure, li2025fast, singh2026privacy, maniveena2024security, maniveena2024security1, shukla2025effective} & 73 \\

& Data generalization & \cite{liu2023post, tian2020research, ma2018architecture, sanchez2018integration, shouqi2019improved, otgonbayar2018k, telikani2023edge, singh2022framework, gheisari2023ppdmit, iwendi2020n, xu2022ppta, zhou2019unlinkable, nukavarapu2022iknight, tabassum2021privacy, imtiaz2021machine, puri2020data, jourdan2018toward, eyeleko2024microaggregation, sangaiah2023privacy, li2025lpps, ma2025personalized, kasula2025federated, alrayes2025privacy} & 23 \\

& Data perturbation & \cite{gheisari2023agile, an2020lopro, zhao2020privacy, alcaide2013anonymous, lu2017lightweight, yan2019privacy, xiong2019locally, humayun2020privacy, tian2021bi, singh2021end, jain2023noise, yassine2015smart, cao2018scrappor, wang2018differentially, qiao2019effective, mahdikhani2019achieving, chen2019internet, zhang2021mpdp, xiao2021analysis, sasada2023oblivious, hindistan2023hybrid, wu2022secure, hamza2020privacy, hou2023block, xiong2022network, chamikara2021privacy, liu2018epic, liu2019dynapro, yan2019location, albouq2020double, zhang2022dprl, aleroud2022privacy, xiong2016randomized, fenelon2023private, wang2023privacy, abi2021fog, zhang2019differentially, fan2021privacy, meshram2023efficient, shivran2023privacy, li2024contract, shi2024data, aghvamipanah2024activity, ni2023ldp, bahbouh2024double, Al-Balasm2024ehenhancing, sun2024low, liu2024multi, yadav2024hybrid, kil2024optimization, abdelraouf2024privacy, villegas2024optimizing, shnain2024privacy, amjath2024renyi, chen2024rae, qiao2025multi, kouekam2025advancing, guduri2023blockchain, kashif2025ai, kumar2025deep, liu2024distributed, jeyakumar2024innovative, zhang2025data, Bharat2025fed, shweta2025fds, gheisari2025enhancing, hu2025efficient, muntather2025federated, wang2025fedmps, jin2025privacy, bi2024achieving, wahida2024adversarial, yao2024privacy, zhao2024design, qashlan2024differential, sahu2024enhanced, mun2024privacy, tan2024privacy, tan2024federated, chen2024private, cao2025hybrid, zheng2025awe, zhang2025privacy, liu2024deep, li2025aldp} & 85 \\

& Data Pseudonymization & \cite{nkenyereye2019towards, al2019misty, suomalainen2016enhancing, huang2016software, rajput2016hierarchical, ji2018efficient, deebak2019authentic, aliev2020matrix, alharthi2021privacy, nicolazzo2021anonymous, zhang2021sport, sezer2023ppfchain, casanova2023maximizing, qi2019time, wang2022blockchain, eddine2021easbf, sun2022secure, sui2017study, al2017seamless, shahzad2021privacy, hindistan2023hybrid, zhang2024privacy, li2024location, mi2024privacy, warrier2024privacy, zeng2024ssg, liu2024accuracy, nazir2025blockchain, guan2024blockchain, mianji2025dynamic, wang2025p3fl, fu2025prise, li2025user, dewangan2023privacy, samanthula2023privacy, yang2024anonymous, desai2023preserving, liu2024privacy, xia2024decentralized, samal2025lightweight, al2023privacy, dos2023dynamic, wang2023conditional, li2023mcpap, parmar2023privacy, liu2023ptap, nair2023privacy, yang2023scalable, su2024efficient, zhou2024heterogeneous, han2024cppa, shen2024combining, ameur2024enhancing, jin2024epaka, yazdinejad2024hybrid, liu2024ipmrss, wang2024lpf, de2024protect, guo2025cloud, wang2025privacy, wang2025dynamic, masood2025dllpm, zhang2025privacy, esmaeelzadeh2024privacy, praveen2023improved, sharma2024enhancing, dayyani2024siot, bai2025using, dou2025message, vijay2025lattice, belgaum2024novel, nagpal2024novel} & 72 \\

& Data Swapping & \cite{yamin2020new} & 1 \\

Data Flow Privacy & Data Sharing protection & \cite{sarwar2022efficient, kong2019privacy, deebak2020smart, hu2022efficient, shen2019privacy, jing2019data, rahman2019blockchain, liu2021fair, wu2023privacy, wang2021privacy, moqurrab2021deep, bose2015not, wang2019privstream, enayati2023location, rivadeneira2023confluence, goyal2024enhancing, gamundani2023scalable, zheng2023smart, rehman2024fedge, sun2024joint, mariappan2024privacy, fatima2025edge, li2025puf, li2024distributed, cao2025privacy, telikani2023edge, moulahi2023blockchain, thota2023cap2m, shen2023efficient, shen2023evolutionary, chen2023federated, wang2023federated, pandey2023privacy, jiang2023privacy, wang2024blockchain, ranjan2024apps, mughal2024secure, regan2024balancing, islam2024differentially, chhetri2024enabling, nath2024lbpv, singla2024privacy, dai2024privacy, abdel2024privacy, zhou2024privacy, lu2024privacy, makhdoom2024privysec, aminifar2024privacy, sarkar2024recurrent, jia2024towards, shang2025data, li2025distributed, li2025blockchain, yu2025blockchain, mahajan2025deep, pinto2025enhancing, zhang2025privacy, chen2025robust, prabha2025towards, bhasha2024data, das2023lightweight, chen2024privacy, cheng2023privacy} & 63 \\

& Multi-path routing & \cite{gheisari2019context, hussain2022improving, sharma2019cooperative} & 3 \\

& Data Privacy Verification & \cite{yankson2023small, zhang2023verifiable, akgun2023privacy, shen2023optimal, zhang2024epri, xu2024lb, liu2024wiretap, hou2025efficient, rafique2023securemed} & 9 \\

& Route path protection & \cite{xu2019edge, shin2017secure, shin2019security, yan2019comprehensive, long2014achieving, radenkovic2016towards, sun2019lightweight, li2020deep, huso2023privacy, hussien2020msclp, mukamanzi2022position, alsadhan2024blockchain, geng2024privacy, sridharan2025lie, xiao2024efficient} & 15 \\

& Collaborative PP & \cite{kwabena2019mscryptonet, qin2019privacy, trivedi2023homomorphic, huang2022improved, putrada2024homomorphic, rahman2024privacy, wang2023privacy, gao2025atlas, nowak2025decentralised, almansour2025privacy, li2023eppsq, duy2023fedchain, hamouda2023ppss, cao2023privacy, sutradhar2024blockchain, alamer2024privacy, yang2024privacy, xing2024privacy, wang2024rflpv, sharma2025blockchain, zhang2025edge, shan2025lightweight, wang2025privacy, ramadan2025secureiot, li2025construction, he2023efficient, anitha2024privacy, gan2025privacy} & 28 \\

Access Control & RBAC & \cite{tao2018multi, mustafa2015dep2sa, saha2019privacy, sharma2023blockchain, xie2025traceability, cai2024using, said2025new, kim2025role} & 8 \\

& ABAC & \cite{zhang2023enabling, belguith2020proud, navdeti2021privacy, li2022privacy, zahra2017fog, cui2018achieving, zheng2018efficient, perez2018lightweight, zhou2021secure, tanveer2021privacy, singh2021efficient, ruan2023policy, singh2025enhancing, tian2023lightweight, niu2024privacy, elhoseny2025rt, garcia2023privacy, li2024smart, xu2023privacy, zhou2024blockchain} & 20 \\

& RBAC and ABAC & \cite{jayaraman2017privacy} & 1 \\

& BBAC & \cite{sasikumar2023blockchain, deebak2022robust, bera2020designing, li2022flexible, ferraro2018distributed, cha2018blockchain, lv2019iot, ma2019privacy, luong2022privacy, gohar2022patient, zhuo2023efficient, li2023three, loukil2021data, manogaran2023token, rahman2018blockchain, safwat2021comparison, acharya2023decentralized, singh2022blockchain, hidayat2023privacy, dongre2025scalable, vangala2022privacy, alsuqaih2023efficient, iftikhar2025blockchain} & 23 \\

& CapBAC & \cite{ren2021privacy, jose2016improving, alshahrani2019secure, geloczi2023unveiling} & 4 \\

& UCON & \cite{cha2018user, gheisari2019eca, chen2019secure, dixit2024privacy} & 4 \\

& ReBAC & \cite{parne2018segb} & 1 \\

& DAC & \cite{liu2020physically, nimmy2021lightweight, jain2023design} & 3 \\

Data Privacy Awareness & - & \cite{pathmabandu2023privacy, zhong2020multi, aloqaily2016multiagent, sendhil2022verifiable, tavakolan2020applying, rivadeneira2023blockchain, alhirabi2023parrot, alhirabi2024designing, windl2024designing, pathmabandu2023privacy, zheng2024ppsfl, choi2024predicting} & 12 \\

Data Isolation & - & \cite{mugarza2019dynamic, kontar2021internet, liang2020fog, radjaa2023federated, shahid2019quantifying, dp2023hierarchical, li2023fedapi, chavhan2024edge, kumar2024trustworthy, sharma2025leveraging, chen2025fog, chien2023enc2, duan2022privacy, zuo2023towards, wang2024lftda, cui2025flav, alam2024blockchain, akter2024spei, pinto2024towards} & 18 \\

Data Minimization & - & \cite{xu2019privacy, tudor2020bes, zhao2022practical, zhou2020top, altowaijri2021privacy, zanddizari2022privacy, sei2022private, al2019fog, salim2022data, gati2020differentially, can2021privacy, xue2020efficient, cui2022boosting, anitha2024federated, revathy2024integrating, isobe2025federated, ali2024federated, jiang2025federated, najm2025pioneering, rajagopal2025pef} & 20
\end{tblr}

\end{table*}

\begin{table*}[htbp]
\caption{Papers on PPT with multiple privacy goals}
\label{tab: multiple goals}
\fontsize{9pt}{9pt}\selectfont
\centering

\begin{tabularx}{\linewidth}{|p{0.49\linewidth}|X|}
\hline
\textbf{Privacy Goals} & \textbf{Paper} \\ \hline

Access control and Data Isolation & \cite{dileep2023preserving} \\ \hline

Anonymity and Data flow protection & \cite{sahlabadi2023gdp, qiu2020electric, yu2025guardgrid, kumari2024lightweight, li2024privacy, zhao2025blockchain, kumar2025connected, wu2023privacy, tang2023efficient, guo2024dbcpca, alamer2023privacy, kashyap2025privacy, zhang2024v2v, ahamad2023two, liu2024bcrs, cheng2024decentralized, de2024protect, han2024privacy, zhang2024privacy, wang2025advancing, dhavamani2024differential, shen2024privacy, basudan2024privacy, verma2024dynamic, chang20232d2ps, antwi2024enhanced, Jaganraja2025enhancing, saeed2025lightweight, yi2024private, sun2025privacy, rajasekar2025privacy, liang2025privacy, xue2025tripartite} \\ \hline

Access control and Data Minimization & \cite{kashif2021bcpripiot, gritti2019privacy} \\ \hline

Anonymity and Data Privacy Awareness & \cite{jagdale2024privacy} \\ \hline

Anonymity and Access control & \cite{qashlan2021privacy, yang2023anonymous, agrawal2022security, psychoula2018deep, chehab2018towards, mohan2023iot, jain2023design, zhang2024towards, xia2025privacy, guduri2023blockchain} \\ \hline

Anonymity and Data isolation & \cite{salim2022perturbation, jiang2023federated, kapila2022federated, xue2018acies, zhu2022data, ngan2022privacy, dhasarathan2023user, wang2024rpifl} \\ \hline

Anonymity and Data Minimization & \cite{kashif2021bcpripiot, gritti2019privacy, kandasamy2024anonymity, suo2023multitiered, villegas2024optimizing, mahmood2025privacy, wang2025achieving, li2024theta} \\ \hline

Data flow protection and Data minimization & \cite{deebak2022ai, agbaje2023privacy, zhang2023two, khan2023blockchain, anitha2023unified} \\ \hline

Data minimization and Data Isolation & \cite{kasyap2021privacy, wang2025lppslf, r2025saoa} \\ \hline

Data Flow Privacy and Access control & \cite{peng2021efficient, wang2025policy, ma2025efficient, arpitha2024hybrid, ahmad2025new, yang2025pecd, liu2024dual} \\ \hline

Data Flow Privacy and Data Isolation & \cite{rahmadika2023blockchain, jayagopalan2023intelligent} \\ \hline

Anonymity, Access Control and Data Flow Privacy & \cite{narkedimilli2025fl, liu2025physically} \\ \hline

Anonymity, Data Isolation and Data Flow Privacy & \cite{alla2023federated} \\ \hline

\end{tabularx}

\end{table*}

\subsection {Technology used in building PETs}

 PETs are tools used to create PPTs in a system. These PETs are built using innovative technologies, such as blockchain, machine learning/deep learning, and post-quantum algorithms, among others, as shown in Figure \ref{fig:PP-in-IoT}. Some of these technologies are explained as follows:

    \subsubsection{Blockchain} Blockchain technology decentralizes the control system, fostering a distributed and peer-to-peer transaction environment among anonymous parties within a digital ledger system \cite{lima2018blockchain} \cite{axon2018privacy}. It enhances privacy in IoT systems through smart contracts, which enable trust transactions that prevent identity disclosure and tracking \cite{iftikhar2021privacy}. Blockchain has emerged as a novel protective layer for the internet and its applications, achieving privacy through the provision of authentication, validation, and tracing mechanisms, which serve as robust safeguards for assets within the IoT system. \cite{lima2018blockchain}.  
    
    \subsubsection {Cryptography} This approach encodes the data to be accessed by only authorized users. An example is Homomorphic encryption, which allows encrypted data to be computed at its encrypted state without decrypting it \cite{li2020deep} \cite{trivedi2023homomorphic}.

    \subsubsection {Machine and Deep Learning (MD/DL)} Both ML and DL are subsets of AI; however, ML enables computers to learn and adapt patterns from given data with little human intervention \cite{badillo2020introduction}. DL is more efficient than ML, as it utilizes artificial multi-layers within an artificial neural network to learn and adapt patterns similar to the human brain \cite{sarker2021deep}. The introduction of AI has allowed the implementation of ML/DL in IoT systems for privacy protection. However, their implementation circles around protecting data in process at the nodes or the cloud (service provider end), as seen in federated learning(FL) applications in IoT privacy protection \cite{can2021privacy, zhao2022practical, kontar2021internet, huang2022improved, wu2023privacy, singh2022framework, sezer2023ppfchain, fang2022privacy, halder2023radio, zhao2020privacy, salim2022perturbation, chamikara2021privacy}.
    
    \subsubsection{Mathematical Model} Some studies have proposed novel mathematical models to protect privacy in the IoT environment. Most of these models center on protecting data in transit during communication among IoT system components. The proposed mathematical model either provides cryptography or access control to the system \cite{gheisari2023agile, alshahrani2019secure, suomalainen2016enhancing, zhang2019new, rodriguez2020cooperative, yankson2023small}.
    
    \subsubsection {Optimization} This method provides enhanced privacy protection in IoT systems by optimizing existing approaches like lightweight encryption methods suitable for IoT environments \cite{xu2019edge, satyanarayana2023comparative, liu2018epic, wang2022blockchain, shin2019security, yan2019location, yan2019comprehensive, xiong2022network, telikani2023edge}. 

     \subsubsection {Post-Quantum Algorithm} This technology is used to enhance privacy in IoT systems by hardening defenses against quantum-based privacy breaches. There are a few publications on the post-quantum use of IoT systems; we identified two of the most relevant papers in this study. The first paper of 2023 by Liu et al. \cite{liu2023post} employs a post-quantum algorithm in industrial IoT (IIoT) to ensure quantum-safe privacy protection. They proposed a lattice-based encryption scheme to ensure the encrypted data is quantum-safe. The second paper by Yi et al. \cite{yi2024private} uses post-quantum cryptography as the security foundation for an IoT-based cryptocurrency (an example of an IoT application in e-commerce). The focus is on enabling IoT devices to participate in cryptocurrency markets and transact securely. This preserves the user's privacy using quantum-resistant encryption algorithms against both classical and quantum adversaries \cite{yi2024private}.

    \subsubsection {Infrastructure-based} These PPTs leverage the infrastructure component in an IoT system \cite{tao2018multi,al2019misty, elmisery2016fog, ma2018architecture, xu2019privacy, gohar2022patient, li2023three}. Some active devices that provide access control are deployed physically or virtually as a Network function (NF) to control who, when, and how subject data is accessed in an IoT system.
    
   \subsubsection{Informed Consent Management} This is the only non-technological approach in our study. It relies on policies and laws governing user data privacy, with focus on consent management and user control in alignment with privacy regulations such as GDPR, Canadian Personal Information Protection and Electronic Documents Act (PIPEDA), California Consumer Privacy Act (CCPA), and Brazilian General Data Protection Act (LGPD), as outlined by regulatory \cite{pathmabandu2023privacy, cha2018user, al2019fog, rivadeneira2023blockchain, pathmabandu2023privacy}. However, other present regulatory frameworks that emphasize user control and informed consent that protect privacy rights, as seen in \cite{pinto2025enhancing}.

  \textbf{Our observations}: Some studies combine these PET building technologies to achieve multiple privacy-preserving goals. Examples are Singh et al. \cite{singh2022framework}, Fang et al. \cite{fang2022privacy}, and Zhao et al. \cite{zhao2020privacy}, all of which use ML and blockchain to provide privacy preservation in IoT. Additionally, Salim et al. \cite{salim2022perturbation} and Hindistan \cite{hindistan2023hybrid} both employ ML and DP to develop PET in IoT systems. Moreover, some authors combine three technologies to achieve privacy preservation in IoT systems. For example, blockchain, ML/DL (especially federated learning), and Cryptography were employed by Khan et al. \cite{khan2023blockchain}, Guduri et al. \cite{guduri2023blockchain}, and Cheng et al. \cite{cheng2024decentralized}. Others, such as Rehman et al. \cite{rehman2024fedge}, employ federated learning, cryptography, and an optimization approach. In contrast, Jagdale et al. \cite{jagdale2024privacy} employ a blockchain, a convolutional neural network, and a post-quantum algorithm to preserve privacy in IoT systems.
    
\subsection{IoT Applications and PPTs}
The literature typically focuses on common IoT applications in transportation, home/building, city, health, and industry. However, IoT applications are expanding into social networks and multimedia, as seen in social IoT, the Internet of Multimedia Things (IoMT), and Artificial Intelligence of Things.

    \subsubsection {Social IoT (S-IoT)} This can be seen in the mutually coordinated social cooperation that exists among IoT devices, with their connection strength depending on interaction levels. These connections are built on trust and privacy \cite{sharma2019cooperative}\cite{al2019misty} \cite{hussain2022improving}. These interactions can leverage platforms like Web 3.0, as seen in social media applications such as SM-IoT \cite{salim2022perturbation} \cite{salim2022data}.
    
    \subsubsection {Internet of Multimedia of Things (IoMT)} This is the smart operation in multimedia systems. Privacy protection in the Internet of Multimedia of Things (IoMT) has recently garnered significant attention from researchers. Various methods have been introduced to protect sensitive data. For example, Elmisery et al. \cite{elmisery2017cognitive} proposed a data mashup service tool for aggregating data from web applications, and Gati et al. \cite{gati2020differentially} proposed a deep private tensor train autoencoder (dPTTAE) technique for handling large volumes of data in smart environments. Additionally, Liang et al. \cite{liang2020fog} proposed a Secure Service Discovery (SSD) approach to protect data communication during the service discovery process, addressing vulnerabilities to eavesdropping and data tampering. Deep learning has emerged as the dominant building technology for PET in IoMT applications, as shown in Figure \ref{fig:PET-Tech}.

   \subsubsection{Artificial Intelligence of Things (AIoT)} The AIoT combines Artificial Intelligence (AI) and IoT in which the Smart device, not only collects information, but understands it, learn this information, and makes decisions solely without human intervention through the help of AI \cite{wang2025acsfl, wang2025privacy}. AI in IoT is typically implemented at edge devices, where raw data is processed locally and differential privacy is applied, adding controlled noise to each device's model before sharing. The device shares the learned data with the central location while leaving the raw data on the device \cite{wang2025acsfl, wang2025privacy}.  Park et al. \cite{park2025lmsa} employ the lightweight key exchange and hashing to achieve strong encryption without exposing sensitive data. In their approach to privacy preservation in AIoT, the authors Ivanovska et al. \cite{ivanovska2025privacy} introduce a comprehensive privacy-by-design approach to implement a PPT directly on the hardware and system architecture preserve the privacy of the data collected by these hardware and an AI processing pipeline \cite{ivanovska2025privacy}. The application of AIoT has cut across IoT applications such as Smart Homes \cite{wang2025privacy}, Smart City \cite{ivanovska2025privacy}, and MIoT \cite{park2025lmsa}.

\textbf{Our observations}: According to Figure \ref{fig:PET-Tech}, the number of papers on PET building technology specific to IoT applications in agriculture, education, and e-commerce is the lowest. In contrast, the fields of Smart Homes, Medical IoT, and Industrial IoT (IIoT) contain a significant number of papers within our study, as illustrated in Figure \ref{fig:PET-Tech}. Additionally, some papers propose PET building technologies that could be applied across all IoT applications. However, this general approach may not be suitable for specific IoT applications, given the differences in architectures, protocols, resource requirements, and communication processes, as noted by Sana et al. \cite{imtiaz2019case}. Moreover{, the introduction of AIoT in IoT environment, gives brain to the IoT devices to learn, reason and make decision without human intervention. However, this may open the door to another aspect of privacy threat that could impact the decisions of these IoT devices.

\subsection {PPT implementation in IoT Application and System Architecture}
PPT implementation in IoT has been proposed across different IoT architectures. These architectures have different layers where PPTs can be deployed. We focus on the implementation of PPTs in IoT computing layers as shown in Figure \ref{fig:computing-layer}.

PPT can be locally deployed within the IoT system at the aggregate (gateway) or at the edge device. Fog and cloud-based architectures are also commonly used today, and most privacy studies have utilized some of these computing layers in IoT architectures to deploy PPTs. There are many vulnerabilities associated with specific computing layers, as some of these layers are managed by third parties, which could result in jurisdictional risks privacy threats \cite{nweke2022linddun}. Additionally, some cloud computing platforms hosted outside the privacy regulations applicable to the location of the IoT system's deployment may not comply with those regulations. This is one of the privacy threats that denies user control over their data, as mentioned in Canadian PIPEDA fair information principle \cite{pipeda-principle}.

\textbf{Our observations:} Some proposed studies on PPTs are deployed across two or more layers of the IoT architectures. Table \ref{table:IoT-Architecturetable} shows the categorization of papers on PPTs and IoT computing layers in which they are deployed. Research on end-to-end PPT deployment in IoT architectures, which span local, fog, edge, and cloud, is limited, as shown in Table \ref{table:IoT-Architecturetable}. Additionally, it was observed that the cloud is the most widely used platform for deploying PPTs in IoT architectures, followed by the edge and then fog. However, many papers did not specify which computing layer they deployed PPT in, as shown in Figure \ref{fig:PPT-implementation layer}. 

\begin{figure}
        \centering
        \includegraphics[width=8cm,height=6cm]{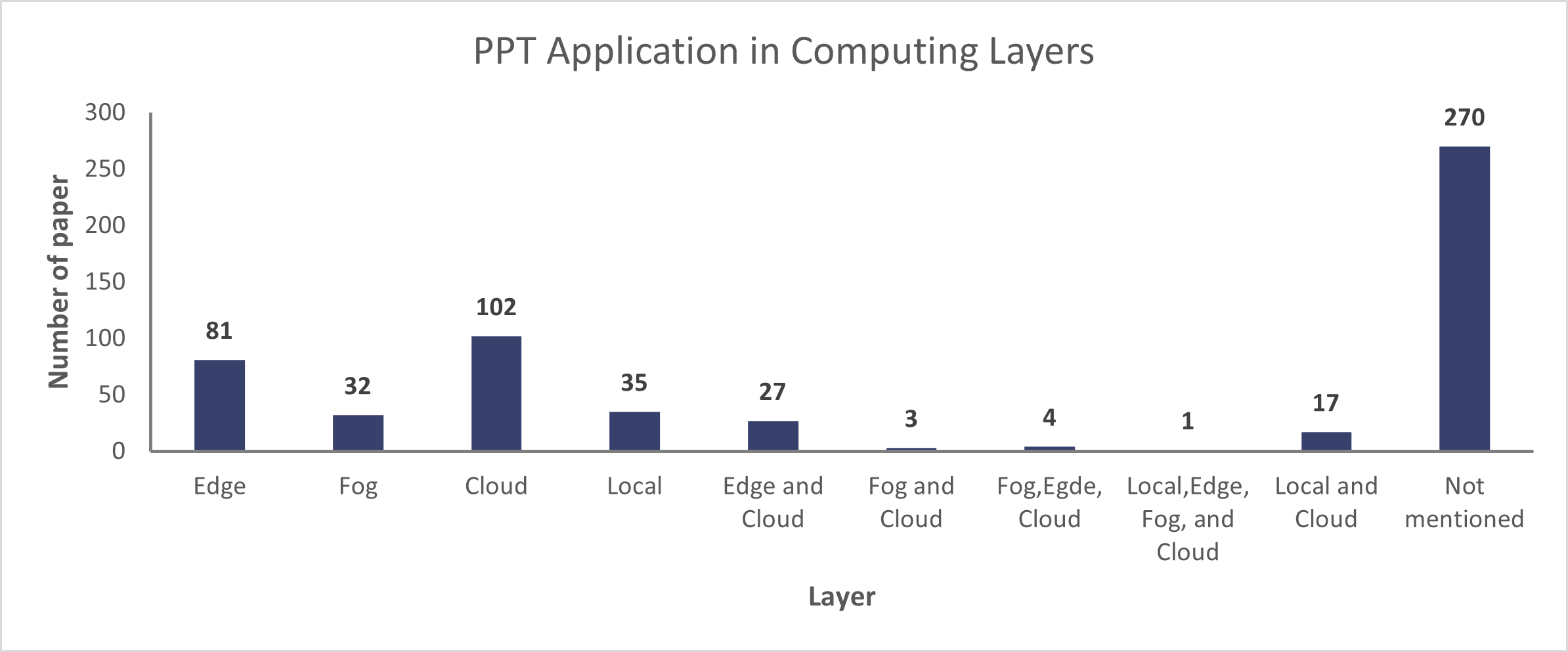}
        \caption{Publications on PPTs implementation at different computing layers in IoT architecture from 2010 to 2025}
        \label{fig:PPT-implementation layer}
\end{figure}
\subsection{Privacy Types provided by PPTs in IoT systems}
\label{privacy-type}

As gathered in this study, the three main privacy types focused on in most papers are user privacy, device privacy, and location privacy. User data privacy protects sensitive information, such as Personally Identifiable Information (PII), Protected Health Information (PHI), or any other data that can reveal a user's identity.
The protection of data stored, processed, or transmitted across devices that could reveal either user or device identity in IoT is called device data privacy. For example, most publications in IIoT primarily focus on the collection, processing, and transmission of device data and data related to industry personnel. Application of PPT to focus on preserving the identity of these production devices could protect not only the device’s identity but also its known vulnerabilities, which malicious threat actors could exploit.
Lastly, location privacy protects IoT users against privacy threats, such as localization and tracking. This threat also applies to mobile IoT devices, including wearable and stationary devices.

Apart from these three privacy types commonly found in IoT systems, some studies have proposed PPTs that simultaneously mitigate both user and device privacy. Meanwhile, some studies have presented the general privacy of the entire IoT system without specifying any particular type of privacy.

\textbf{Our observations}: This study presents different privacy types proposed from the literature on PPT implementation in IoT systems. User data privacy has the highest number of papers focusing on user data protection. Few papers proposed PPT to protect device data and location privacy, as shown in Table \ref{table:Privacy Types} and Figure \ref{fig:Privacy-Types}. This answers our research question RQ6.

\begin{figure}
        \centering
    \includegraphics[width=8cm,height=6cm]{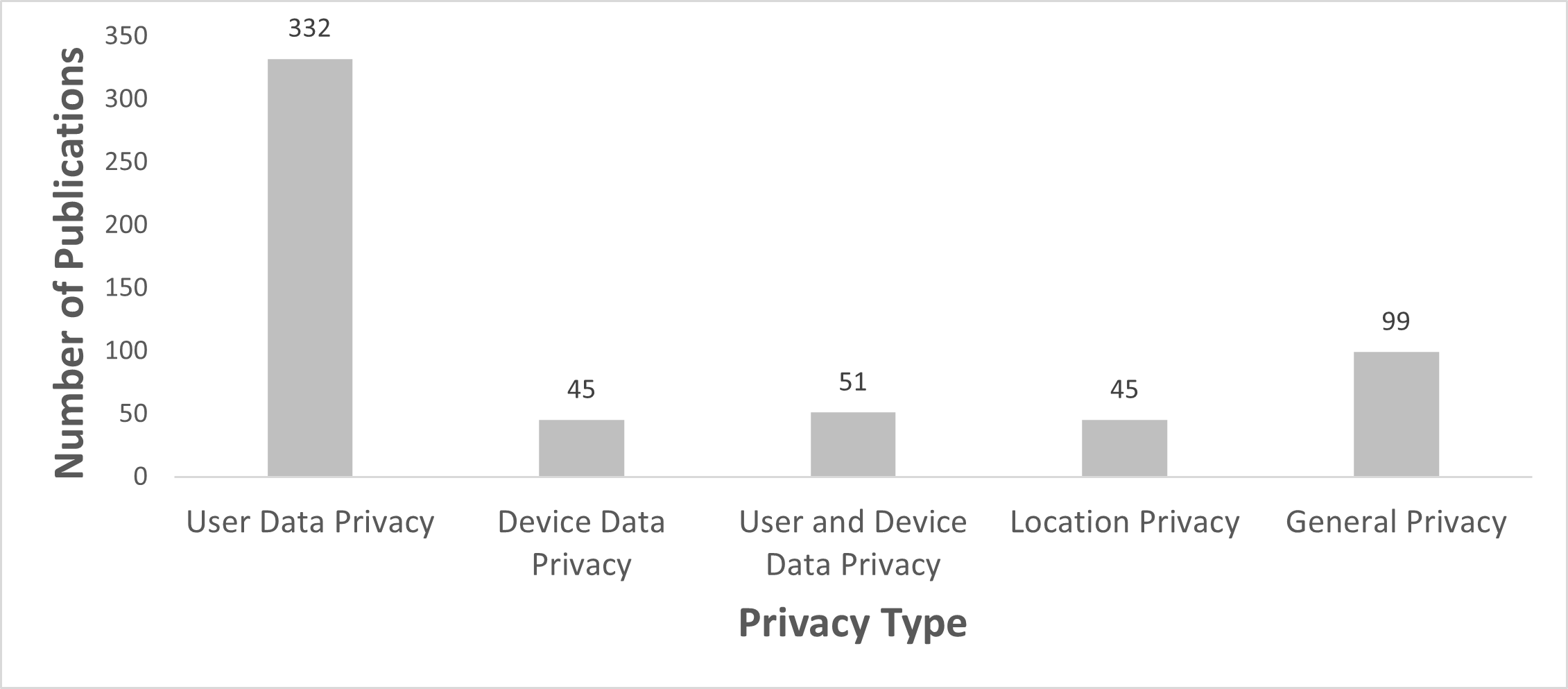}
        \caption{Papers on Privacy protection types in IoT systems from 2010 to 2025}
        \label{fig:Privacy-Types}
\end{figure}

We observed that device data privacy, as proposed by most studies, still focuses on user data being stored, processed, or transmitted across IoT devices. However, any successful threat to the device's identity privacy could compromise both the user's and the device's data stored in the device. Therefore, it is essential to protect device identity and sensitive device data, such as IP addresses, MAC addresses, operating systems, and serial numbers, among others, that could reveal device identity. These sensitive data associated with IoT devices could be linked with other data to identify these devices in the IoT system.

\section{Step 5: Collating, Summarizing, and Reporting Results that Address Research Questions}
\label{section:Summary}

Using the PRISMA-ScR guidelines for selecting relevant papers, our analysis reveals a steady increase in publications on PPTs from 2013 to 2019, with 2019 having the highest number of publications, as shown in Figure \ref{fig:year-of-publications}. There was a noticeable decline in 2020, probably due to the global pandemic. However, publication rates rebounded and continued to rise from 2021 to the end of 2025, when the paper collection was concluded. The following subsections present results that address the research questions outlined in section \ref{section:RQ}.

\subsection{RQ1: What technologies are prominently used to develop PETs in IoT systems?}
    
To address RQ1, this study analyzed seven technologies used to build privacy-enhancing technologies (PETs), including blockchain, Cryptography, Machine Learning and Deep Learning, Mathematical Models, Optimization, Infrastructure-Based techniques, and post-quantum algorithms. It has also been revealed that cryptography is the most prevalent technology used for privacy preservation in IoT applications, followed closely by blockchain. As observed in this study, the two publications on post-quantum algorithms are primarily applied to Industrial IoT (IIoT) and e-commerce IoT, respectively, indicating their pioneering role and potential for future research in IIoT privacy. However, informed consent management is the only non-technical approach to preserving privacy, as our study suggests.

\subsection{RQ2:  What are the dominant PPT’s privacy goals in IoT
applications?}
Our study's research question (RQ2) examines the primary objectives of PPTs across various IoT applications. Our analysis identifies various PPT goals, including anonymity, data flow privacy, access control, privacy awareness, data isolation, and data minimization, as shown in Table \ref{tab: PPTGoal table}. Among these goals, anonymity, particularly through methods such as data masking, data perturbation, and pseudonymization, emerges as the most prevalent objective in the literature, with a significant percentage of publications focusing on these PPT approaches.}

\subsection{RQ3: What are the commonly known and recent IoT
applications added to the literature?} 
IoT applications commonly span cities, transportation, health, and agriculture. However, new application areas like Social IoT(S-IoT) Internet Multimedia of Things (IoMT) and Artificial Intelligence of Things (AIoT) are gaining traction as shown in Figure \ref{fig:PET-Tech}. These new platforms directly interact with users and could be vulnerable to privacy threats due to their extensive internet exposure.


\onecolumn
\begin{landscape}
\scriptsize  
\setlength{\tabcolsep}{3pt}
\renewcommand{\arraystretch}{1.1}
\hspace{-4cm}
\begin{longtable}{|p{1.8cm}|p{3.0cm}|p{1.0cm}|p{3.0cm}|p{1.5cm}|p{1.2cm}|p{1.0cm}|p{1.3cm}|p{1.3cm}|p{1.0cm}|p{5.0cm}|}

\caption{Papers on PPTs and Layer of Implementation in IoT architecture from 2010 to  2025.} 
\label{table:IoT-Architecturetable} \\
\hline
\textbf{IoT Application} & \textbf{Edge} & \textbf{Fog} & \textbf{Cloud} & \textbf{Local} & \textbf{Edge \& Cloud} & \textbf{Fog \& Cloud} & \textbf{Fog, Edge \& Cloud} & \textbf{Local, Edge, Fog \& Cloud} & \textbf{Local \& Cloud} & \textbf{Not Mentioned} \\
\hline
\endfirsthead

\multicolumn{11}{c}{\scriptsize\textit{Table \ref{table:IoT-Architecturetable} continued}} \\
\hline
\textbf{IoT Application} & \textbf{Edge} & \textbf{Fog} & \textbf{Cloud} & \textbf{Local} & \textbf{Edge \& Cloud} & \textbf{Fog \& Cloud} & \textbf{Fog, Edge \& Cloud} & \textbf{Local, Edge, Fog \& Cloud} & \textbf{Local \& Cloud} & \textbf{Not Mentioned} \\
\hline
\endhead

\hline
\endfoot

\hline
\endlastfoot

\textbf{Smart Home} & 
\cite{zhao2020privacy}, \cite{rahman2018blockchain}, \cite{safwat2021comparison}, \cite{waheed2023privacy}, \cite{gritti2019privacy}, \cite{chehab2018towards}, \cite{mariappan2024privacy}, \cite{qashlan2024differential} & 
\cite{alshahrani2019secure}, \cite{puri2020data} & 
\cite{pathmabandu2023privacy}, \cite{tao2018multi}, \cite{shin2017secure}, \cite{shin2019security}, \cite{xu2019privacy}, \cite{psychoula2018deep} & 
\cite{gamundani2023scalable}, \cite{cheng2023privacy}, \cite{aghvamipanah2024activity} & 
\cite{qashlan2021privacy} & 
--- & --- & --- & 
\cite{meshram2023efficient} & 
\cite{jose2016improving}, \cite{perez2018lightweight}, \cite{yu2021lightweight}, \cite{tian2021bi}, \cite{nimmy2021lightweight}, \cite{chintan}, \cite{putrada2024homomorphic}, \cite{2025rehan}, \cite{zhao2024design}, \cite{wang2025p3fl}, \cite{dos2023dynamic} \\ 
\hline

\textbf{Industrial IoT} & 
\cite{wu2023privacy}, \cite{hamouda2023ppss}, \cite{odeh2025secure}, \cite{hindistan2023hybrid} & 
\cite{li2022privacy}, \cite{han2024privacy} & 
\cite{deebak2022robust}, \cite{li2020deep}, \cite{he2023efficient} & 
\cite{liu2023post}, \cite{sui2017study}, \cite{hu2025efficient}, \cite{wang2025fedmps} & 
\cite{deebak2022ai}, \cite{saeed2025lightweight} & 
\cite{wang2025dynamic} & 
\cite{humayun2020privacy} & 
--- & --- & 
\cite{sasikumar2023blockchain}, \cite{halder2023radio}, \cite{liu2021fair}, \cite{yang2025pecd}, \cite{duy2023fedchain}, \cite{ramadan2025secureiot}, \cite{pandey2023privacy}, \cite{regan2024balancing}, \cite{islam2024differentially}, \cite{sarkar2024recurrent}, \cite{eyeleko2024microaggregation}, \cite{li2025aeppfl}, \cite{mahendran2025prism}, \cite{kumar2025deep} \\
\hline

\textbf{Smart Grid} & 
\cite{chen2019internet}, \cite{bose2015not}, \cite{nowak2025decentralised}, \cite{zhou2024heterogeneous} & 
\cite{zhang2024epri} & 
\cite{singh2021end}, \cite{mustafa2015dep2sa}, \cite{shahzad2021privacy}, \cite{samanthula2023privacy} & 
\cite{chen2023federated}, \cite{wang2023federated}, \cite{li2023mcpap} & 
\cite{wang2021privacy}, \cite{li2024privacy}, \cite{shen2023evolutionary}, \cite{akgun2023privacy} & 
--- & --- & --- & 
\cite{shivran2023privacy} & 
\cite{tudor2020bes}, \cite{deng2022secure}, \cite{yassine2015smart}, \cite{yu2025guardgrid}, \cite{chang20232d2ps}, \cite{li2023eppsq}, \cite{dai2024privacy}, \cite{chang2023practical}, \cite{liu2024distributed}, \cite{zhang2025data}, \cite{jin2025privacy}, \cite{zhang2025privacy}, \cite{nazir2025blockchain}, \cite{li2025user} \\ 
\hline

\textbf{Medical IoT} & 
\cite{trivedi2023homomorphic}, \cite{tabassum2021privacy}, \cite{dileep2023preserving}, \cite{imtiaz2021machine}, \cite{jiang2023federated}, \cite{jourdan2018toward}, \cite{ngan2022privacy}, \cite{kumari2024lightweight}, \cite{cao2023privacy}, \cite{aminifar2024privacy}, \cite{shweta2025fds}, \cite{nair2023privacy} & 
\cite{elmisery2016fog}, \cite{saha2019privacy}, \cite{kasyap2021privacy}, \cite{mughal2024secure} & 
\cite{wang2021forward}, \cite{singh2022framework}, \cite{guo2020new}, \cite{zhang2023enabling}, \cite{zheng2018efficient}, \cite{tan2019secure}, \cite{agrawal2022security}, \cite{xu2022cryptoanalysis}, \cite{huang2022improved}, \cite{zhang2021mpdp}, \cite{ming2020efficient}, \cite{zhang2021sport}, \cite{luong2022privacy}, \cite{gohar2022patient}, \cite{fang2020privacy}, \cite{sharma2023blockchain}, \cite{wang2022panda}, \cite{singh2021efficient}, \cite{ma2025efficient}, \cite{liu2024dual}, \cite{guduri2023blockchain}, \cite{jayagopalan2023intelligent}, \cite{jagdale2024privacy}, \cite{alsadhan2024blockchain}, \cite{rafique2023securemed}, \cite{al2023privacy} & 
\cite{deebak2019authentic}, \cite{r2025saoa}, \cite{wang2024rflpv}, \cite{li2024distributed} & 
\cite{sahlabadi2023gdp} & 
--- & 
\cite{sutradhar2024blockchain} & 
--- & 
\cite{liu2025physically}, \cite{samal2025lightweight}, \cite{guo2025cloud} & 
\cite{iwendi2020n}, \cite{hamza2020privacy}, \cite{moqurrab2021deep}, \cite{zhang2018tolerating}, \cite{arpitha2024hybrid}, \cite{kashyap2025privacy}, \cite{khan2023blockchain}, \cite{kandasamy2024anonymity}, \cite{dhasarathan2023user}, \cite{rahmadika2023blockchain}, \cite{shan2025lightweight}, \cite{wang2025privacy}, \cite{moulahi2023blockchain}, \cite{thota2023cap2m}, \cite{shen2023efficient}, \cite{zhang2025privacy}, \cite{chen2025robust}, \cite{prabha2025towards}, \cite{bhasha2024data}, \cite{das2023lightweight}, \cite{sangaiah2023privacy}, \cite{amjath2024renyi}, \cite{chen2024rae}, \cite{kouekam2025advancing}, \cite{guan2024blockchain}, \cite{liu2024ipmrss}, \cite{esmaeelzadeh2024privacy}, \cite{praveen2023improved}, \cite{sharma2024enhancing}, \cite{vijay2025lattice}, \cite{belgaum2024novel} \\
\hline

\textbf{Social IoT} & 
\cite{sharma2019cooperative}, \cite{maniveena2024security}, \cite{maniveena2024security1}, \cite{shukla2025effective} & 
--- & 
\cite{al2019misty}, \cite{zhang2019differentially} & 
--- & --- & --- & --- & --- & --- & 
\cite{salim2022data}, \cite{salim2022perturbation}, \cite{hussain2022improving}, \cite{dong2025dummy}, \cite{dayyani2024siot} \\
\hline

\textbf{IoV} & 
\cite{xu2019edge}, \cite{agbaje2023privacy}, \cite{yao2024privacy}, \cite{mianji2025dynamic} & 
\cite{nkenyereye2019towards}, \cite{navdeti2021privacy}, \cite{xia2021effective}, \cite{albouq2020double}, \cite{eddine2021easbf} & 
\cite{chen2023bcgs}, \cite{deebak2020smart}, \cite{bera2020designing}, \cite{huang2016software}, \cite{aloqaily2016multiagent}, \cite{chen2019secure}, \cite{zhuo2023efficient}, \cite{liu2023situ}, \cite{yang2024privacy}, \cite{li2025blockchain}, \cite{Al-Balasm2024ehenhancing} & 
\cite{fenelon2023private}, \cite{zhao2025blockchain}, \cite{liu2024wiretap}, \cite{shi2024data}, \cite{wang2023conditional}, \cite{yang2023scalable}, \cite{bai2025using} & 
\cite{belguith2020proud}, \cite{qiu2020electric}, \cite{singh2022blockchain}, \cite{wang2025advancing}, \cite{shen2024privacy}, \cite{li2025distributed} & 
\cite{kapila2022federated} & 
--- & --- & 
\cite{tang2023efficient}, \cite{almansour2025privacy}, \cite{cao2025privacy}, \cite{hou2025efficient} & 
\cite{kong2019privacy}, \cite{rajput2016hierarchical}, \cite{sui2017study}, \cite{aliev2020matrix}, \cite{ahmed2020anonymous}, \cite{zhou2020top}, \cite{alharthi2021privacy}, \cite{an2020lopro}, \cite{manogaran2023token}, \cite{kumar2025connected}, \cite{guo2024dbcpca}, \cite{zhang2024v2v}, \cite{verma2024dynamic}, \cite{sun2025privacy}, \cite{rajasekar2025privacy}, \cite{xue2025tripartite}, \cite{li2024theta}, \cite{xiao2024efficient}, \cite{sun2024joint}, \cite{jiang2023privacy}, \cite{wang2024blockchain}, \cite{nath2024lbpv}, \cite{zhou2024privacy}, \cite{shang2025data}, \cite{xu2024lb}, \cite{li2025lpps}, \cite{ma2025personalized}, \cite{jibril2024semantic}, \cite{shen2025privacy}, \cite{li2025fast}, \cite{sahu2024enhanced}, \cite{mun2024privacy}, \cite{tan2024privacy}, \cite{chen2024private}, \cite{liu2024deep}, \cite{mi2024privacy}, \cite{fu2025prise}, \cite{parmar2023privacy}, \cite{liu2023ptap}, \cite{su2024efficient}, \cite{han2024cppa}, \cite{shen2024combining}, \cite{ameur2024enhancing}, \cite{jin2024epaka}, \cite{wang2024lpf}, \cite{wang2025privacy}, \cite{masood2025dllpm}, \cite{dou2025message} \\ 
\hline

\textbf{Wearable IoT} & 
\cite{rehman2024fedge} & 
--- & 
\cite{ji2018efficient}, \cite{zanddizari2022privacy}, \cite{aleroud2022privacy}, \cite{mohan2023iot}, \cite{bigelli2024privacy} & 
\cite{casanova2023maximizing}, \cite{marchioro2023practical} & 
--- & --- & --- & --- & --- & 
\cite{zhang2019new}, \cite{can2021privacy}, \cite{deepak2024privacy}, \cite{warrier2024privacy} \\
\hline

\textbf{Smart City} & 
\cite{xie2022improved}, \cite{gheisari2019eca}, \cite{rahman2019blockchain}, \cite{Jaganraja2025enhancing}, \cite{Bharat2025fed}, \cite{gheisari2025enhancing}, \cite{wahida2024adversarial} & 
\cite{yamin2020new}, \cite{abi2021fog}, \cite{abdel2024privacy}, \cite{bahbouh2024double} & 
\cite{gheisari2023agile}, \cite{gope2018lightweight}, \cite{mugarza2019dynamic}, \cite{zhao2025reversible} & 
--- & 
\cite{gheisari2019context} & 
--- & --- & --- & 
\cite{liu2018epic}, \cite{ferraro2018distributed}, \cite{narkedimilli2025fl}, \cite{wang2023privacy} & 
\cite{shen2019privacy}, \cite{alcaide2013anonymous}, \cite{suomalainen2016enhancing}, \cite{de2024protect}, \cite{alla2023federated}, \cite{rahman2024privacy}, \cite{gan2025privacy}, \cite{rivadeneira2023confluence}, \cite{li2025puf}, \cite{jaganraja2025agile}, \cite{singh2026privacy}, \cite{tan2024federated}, \cite{dewangan2023privacy} \\ 
\hline

\textbf{IoMT} & 
--- & 
\cite{liang2020fog} & 
\cite{elmisery2017cognitive} & 
--- & --- & --- & --- & --- & --- & 
\cite{gati2020differentially} \\
\hline

\textbf{Agriculture} & 
--- & --- & 
\cite{yan2021scalable} & 
--- & --- & --- & --- & --- & --- & 
\cite{sharma2025blockchain}, \cite{zheng2023smart}, \cite{makhdoom2024privysec} \\
\hline

\textbf{Education} & 
--- & --- & 
\cite{xiao2021analysis} & 
--- & --- & --- & --- & --- & --- & 
--- \\
\hline

\textbf{E-commerce} & 
--- & --- & 
\cite{hua2024international} & 
--- & --- & --- & --- & --- & --- & 
\cite{chen2019light}, \cite{yi2024private} \\
\hline

\textbf{General IoT} & 
\cite{wu2022secure}, \cite{satyanarayana2023comparative}, \cite{huso2023privacy}, \cite{zhong2020multi}, \cite{fang2022privacy}, \cite{cui2018achieving}, \cite{wu2019lightweight}, \cite{qiao2019effective}, \cite{yan2019privacy}, \cite{jing2019data}, \cite{zhou2021secure}, \cite{zhang2022dprl}, \cite{telikani2023edge}, \cite{wang2022blockchain}, \cite{wang2019privstream}, \cite{fenelon2023private}, \cite{wang2023privacy}, \cite{fan2021privacy}, \cite{kashif2021bcpripiot}, \cite{xue2020efficient}, \cite{xue2018acies}, \cite{shahid2019quantifying}, \cite{yang2023privacy}, \cite{zhu2022data}, \cite{zhang2025edge}, \cite{fatima2025edge}, \cite{shen2023optimal}, \cite{wang2025privacy}, \cite{yadav2024hybrid}, \cite{bi2024achieving}, \cite{liu2024privacy} & 
\cite{sarwar2022efficient}, \cite{zahra2017fog}, \cite{lu2017lightweight}, \cite{cao2018scrappor}, \cite{mahdikhani2019achieving}, \cite{al2019fog}, \cite{sezer2023ppfchain}, \cite{sendhil2022verifiable}, \cite{radjaa2023federated}, \cite{jain2023design}, \cite{antwi2024enhanced} & 
\cite{gheisari2023ppdmit}, \cite{jayaraman2017privacy}, \cite{nasiraee2020anonymous}, \cite{ren2021privacy}, \cite{yin2022novel}, \cite{yan2017context}, \cite{al2017seamless}, \cite{badsha2018designing}, \cite{wang2018differentially}, \cite{zhang2019provable}, \cite{zhu2019secure}, \cite{yan2019comprehensive}, \cite{lv2019iot}, \cite{kwabena2019mscryptonet}, \cite{qin2019privacy}, \cite{shin2020privacy}, \cite{altowaijri2021privacy}, \cite{yankson2023small}, \cite{zhang2023verifiable}, \cite{chaudhry2021rotating}, \cite{wang2025policy}, \cite{liang2025privacy}, \cite{alamer2024privacy}, \cite{goyal2024enhancing}, \cite{li2024contract} & 
\cite{hu2022efficient}, \cite{zhao2022practical}, \cite{cha2018user}, \cite{peng2021efficient}, \cite{hussien2020msclp}, \cite{tavakolan2020applying}, \cite{cheng2024decentralized}, \cite{basudan2024privacy}, \cite{wang2025lppslf}, \cite{ni2023ldp}, \cite{kil2024optimization}, \cite{muntather2025federated} & 
\cite{wang2018privacy}, \cite{kontar2021internet}, \cite{baccour2021rl}, \cite{alamer2023privacy}, \cite{zhang2024privacy}, \cite{xia2025privacy}, \cite{anitha2023unified}, \cite{wang2024rpifl}, \cite{xia2024decentralized} & 
\cite{ma2019privacy} & 
\cite{chhetri2024enabling} & 
\cite{chamikara2021privacy}, \cite{zhang2023two}, \cite{gao2025atlas}, \cite{kashif2025ai} & 
\cite{radenkovic2016towards} & 
\cite{xu2020aggregate}, \cite{liu2023improvement}, \cite{xu2022ppta}, \cite{li2022flexible}, \cite{tian2020research}, \cite{long2014achieving}, \cite{yang2018density}, \cite{wang2018blockchain}, \cite{ma2018architecture}, \cite{cha2018blockchain}, \cite{parne2018segb}, \cite{sanchez2018integration}, \cite{qiu2019efficient}, \cite{shouqi2019improved}, \cite{yan2019location}, \cite{meng2019steganography}, \cite{sun2019lightweight}, \cite{zhou2019unlinkable}, \cite{xiong2019locally}, \cite{liu2020physically}, \cite{rodriguez2020cooperative}, \cite{wan2020internet}, \cite{alzahrani2021provable}, \cite{pistono2021cryptosystem}, \cite{al2022counterfeit}, \cite{nukavarapu2022iknight}, \cite{sei2022private}, \cite{sasada2023oblivious}, \cite{otgonbayar2018k}, \cite{qi2019time}, \cite{hou2023block}, \cite{xiong2022network}, \cite{yang2023anonymous}, \cite{sun2022secure}, \cite{li2022data}, \cite{li2023three}, \cite{loukil2021data}, \cite{jain2023noise}, \cite{sundarakantham2023hybrid}, \cite{acharya2023decentralized}, \cite{xiong2016randomized}, \cite{zhu2022data}, \cite{ahmad2025new}, \cite{wu2023privacy}, \cite{ahamad2023two}, \cite{liu2024bcrs}, \cite{dhavamani2024differential}, \cite{zhang2024towards}, \cite{suo2023multitiered}, \cite{villegas2024optimizing}, \cite{mahmood2025privacy}, \cite{wang2025achieving}, \cite{xing2024privacy}, \cite{li2025construction}, \cite{anitha2024privacy}, \cite{mukamanzi2022position}, \cite{geng2024privacy}, \cite{sridharan2025lie}, \cite{enayati2023location}, \cite{ranjan2024apps}, \cite{singla2024privacy}, \cite{lu2024privacy}, \cite{jia2024towards}, \cite{yu2025blockchain}, \cite{mahajan2025deep}, \cite{pinto2025enhancing}, \cite{chen2024privacy}, \cite{kasula2025federated}, \cite{alrayes2025privacy}, \cite{karthikeyan2024privacy}, \cite{xu2025privacy}, \cite{yenugula2025privacy}, \cite{sun2024low}, \cite{liu2024multi}, \cite{abdelraouf2024privacy}, \cite{shnain2024privacy}, \cite{qiao2025multi}, \cite{jeyakumar2024innovative}, \cite{cao2025hybrid}, \cite{zheng2025awe}, \cite{li2025aldp}, \cite{zhang2024privacy}, \cite{li2024location}, \cite{zeng2024ssg}, \cite{liu2024accuracy}, \cite{yang2024anonymous}, \cite{desai2023preserving}, \cite{yazdinejad2024hybrid}, \cite{nagpal2024novel} \\
\hline

\end{longtable}
\end{landscape}

\twocolumn

\subsection{RQ4: What computing layer in IoT architectures are PPTs mostly implemented within IoT systems, and which IoT application has PPT been most implemented in the literature?} 
IoT applications have increased recently, with medical IoT having the highest PPTs implementation as seen in Figure \ref{fig:PET-Tech}. This shows the importance of privacy in medical IoT systems. This level of importance is a result of the sensitive medical data transactions involved in the MIoT system, which contains data such as Personally Identifiable Information (PII) and Protected Health Information (PHI). In the event of a data breach, this sensitive medical data can reveal an individual's identity in relation to their medical condition. The next prominent IoT application is the IoV, which has a greater number of publications on PPT in recent years, as seen in \autoref{fig:PET-Tech}. This growth may result from the rise in the number of electric vehicles and driverless cars used in shared rides. Some of these vehicles could reveal the locations of their passengers or take the wrong route in the event of a privacy breach that disrupts their operations. Furthermore, our observation shows that PPTs are more implemented around the cloud layer in the IoT architecture in most IoT applications compared to other computing layers, as shown in Table \ref{table:IoT-Architecturetable}.

\begin{sidewaysfigure*}
\includegraphics[width=23.5cm,height=12cm]{PPT-and-IoT-applications.png}
\caption{Papers on PPT implementation in IoT applications and Technology used to build PETs from 2010 to 2025}
\label{fig:PET-Tech}
\end{sidewaysfigure*}

\begin{table*}[htbp]
\caption{Publications on privacy protection types in IoT systems from 2010 to 2025.}
\label{table:Privacy Types}
\fontsize{8pt}{10pt}\selectfont
\centering
\begin{tabularx}{\linewidth}{|p{0.15\linewidth}|c|X|}
\hline 
\textbf{Privacy Type} & \textbf{Number of papers} & \textbf{Papers}\\ 
\hline 

\textbf{User Data Privacy} & 332 & \cite{gheisari2023agile, deng2022secure, xu2022ppta, li2022flexible, xie2022improved, yin2022novel, zhao2020privacy, mustafa2015dep2sa, yassine2015smart, suomalainen2016enhancing, aloqaily2016multiagent, radenkovic2016towards, jose2016improving, rajput2016hierarchical, elmisery2016fog, zahra2017fog, sui2017study, yan2017context, lu2017lightweight, al2017seamless, shin2017secure, rahman2018blockchain, ji2018efficient, cao2018scrappor, zhang2018tolerating, zheng2018efficient, cha2018user, wang2018blockchain, ma2018architecture, perez2018lightweight, cha2018blockchain, wang2018differentially, elmisery2017cognitive, wu2019lightweight, qiu2019efficient, liu2019dynapro, shouqi2019improved, tan2019secure, shin2019security, deebak2019authentic, zhu2019secure, jing2019data, meng2019steganography, xu2019privacy, saha2019privacy, mugarza2019dynamic, ma2019privacy, chen2019light, zhou2019unlinkable, chen2019secure, yamin2020new, aliev2020matrix, liu2020physically, rodriguez2020cooperative, ming2020efficient, ahmed2020anonymous, zhou2020top, kontar2021internet, yu2021lightweight, qashlan2021privacy, alharthi2021privacy, xiao2021analysis, nicolazzo2021anonymous, tian2021bi, tabassum2021privacy, luong2022privacy, nukavarapu2022iknight, gohar2022patient, agrawal2022security, xu2022cryptoanalysis, zanddizari2022privacy, sei2022private, nimmy2021lightweight, zhuo2023efficient, sasada2023oblivious, hindistan2023hybrid, liu2023situ, yankson2023small, otgonbayar2018k, qi2019time, huang2022improved, zhang2023verifiable, hamza2020privacy, salim2022data, liu2021fair, navdeti2021privacy, wang2022blockchain, wu2022secure, eddine2021easbf, alcaide2013anonymous, jain2023noise, kasyap2021privacy, fang2022privacy, gati2020differentially, salim2022perturbation, xiong2022network, wang2021privacy, yan2021scalable, wu2023privacy, yang2023anonymous, sharma2023blockchain, can2021privacy, loukil2021data, chaudhry2021rotating, moqurrab2021deep, peng2021efficient, manogaran2023token, zhang2021mpdp, humayun2020privacy, nasiraee2020anonymous, ren2021privacy, huso2023privacy, chamikara2021privacy, zhong2020multi, fang2022privacy, liu2023improvement, pathmabandu2023privacy, li2022privacy, humayun2020privacy, deebak2022ai, singh2021end, wang2021forward, singh2022framework, guo2020new, zhang2023enabling, iwendi2020n, wang2022panda, sharma2019cooperative, belguith2020proud, chen2023bcgs, gheisari2019eca, sarwar2022efficient, jayaraman2017privacy, xu2020aggregate, gheisari2023ppdmit, sundarakantham2023hybrid, safwat2021comparison, psychoula2018deep, bose2015not, chehab2018towards, wang2019privstream, wang2023privacy, fan2021privacy, waheed2023privacy, aleroud2022privacy, dileep2023preserving, imtiaz2021machine, sahlabadi2023gdp, abi2021fog, qiu2020electric, zhang2019differentially, puri2020data, singh2022blockchain, kashif2021bcpripiot, tavakolan2020applying, singh2021efficient, xue2020efficient, xue2018acies, xiong2016randomized, jourdan2018toward, gritti2019privacy, yang2023privacy, zhu2022data, ngan2022privacy, wang2025policy, ma2025efficient, arpitha2024hybrid, ahmad2025new, liu2024dual, yu2025guardgrid, kumari2024lightweight, li2024privacy, wu2023privacy, kashyap2025privacy, zhang2024v2v, ahamad2023two, liu2024bcrs, cheng2024decentralized, shen2024privacy, chang20232d2ps, sun2025privacy, xue2025tripartite, mohan2023iot, guduri2023blockchain, zhang2023two, kandasamy2024anonymity, villegas2024optimizing, li2024theta, dhasarathan2023user, wang2024rpifl, wang2025lppslf, r2025saoa, rahmadika2023blockchain, jayagopalan2023intelligent, narkedimilli2025fl, liu2025physically, jagdale2024privacy, putrada2024homomorphic, rahman2024privacy, gao2025atlas, nowak2025decentralised, li2023eppsq, cao2023privacy, sutradhar2024blockchain, yang2024privacy, xing2024privacy, wang2024rflpv, wang2025privacy, anitha2024privacy, alsadhan2024blockchain, rivadeneira2023confluence, rehman2024fedge, sun2024joint, mariappan2024privacy, fatima2025edge, li2025puf, li2024distributed, moulahi2023blockchain, thota2023cap2m, shen2023efficient, shen2023evolutionary, pandey2023privacy, jiang2023privacy, mughal2024secure, chhetri2024enabling, zhou2024privacy, lu2024privacy, makhdoom2024privysec, aminifar2024privacy, jia2024towards, shang2025data, yu2025blockchain, pinto2025enhancing, zhang2025privacy, chen2025robust, prabha2025towards, bhasha2024data, das2023lightweight, chen2024privacy, cheng2023privacy, shen2023optimal, zhang2024epri, rafique2023securemed, sangaiah2023privacy, chang2023practical, marchioro2023practical, bigelli2024privacy, deepak2024privacy, wang2025privacy, zhao2025reversible, li2025fast, maniveena2024security, shukla2025effective, meshram2023efficient, shivran2023privacy, shi2024data, aghvamipanah2024activity, ni2023ldp, abdelraouf2024privacy, amjath2024renyi, chen2024rae, kouekam2025advancing, guduri2023blockchain, kashif2025ai, zhang2025data, shweta2025fds, jin2025privacy, bi2024achieving, wahida2024adversarial, yao2024privacy, zhao2024design, qashlan2024differential, zhang2024privacy, warrier2024privacy, zeng2024ssg, liu2024accuracy, nazir2025blockchain, guan2024blockchain, mianji2025dynamic, li2025user, samanthula2023privacy, yang2024anonymous, desai2023preserving, samal2025lightweight, al2023privacy, dos2023dynamic, li2023mcpap, nair2023privacy, zhou2024heterogeneous, liu2024ipmrss, guo2025cloud, zhang2025privacy, esmaeelzadeh2024privacy, praveen2023improved, sharma2024enhancing, dayyani2024siot, bai2025using, vijay2025lattice, belgaum2024novel}\\ 
\hline 

\textbf{Device Data Privacy} & 45 & \cite{zhao2022practical, tian2020research, wang2018privacy, parne2018segb, mahdikhani2019achieving, sun2019lightweight, alzahrani2021provable, hou2023block, satyanarayana2023comparative, hu2022efficient, chintan, liu2023post, halder2023radio, han2024privacy, basudan2024privacy, saeed2025lightweight, rajasekar2025privacy, duy2023fedchain, hamouda2023ppss, alamer2024privacy, ramadan2025secureiot, he2023efficient, geng2024privacy, islam2024differentially, dai2024privacy, sarkar2024recurrent, liu2024wiretap, eyeleko2024microaggregation, shen2025privacy, mahendran2025prism, odeh2025secure, kumar2025deep, hu2025efficient, wang2025fedmps, tan2024federated, chen2024private, su2024efficient, wang2025dynamic, dou2025message, nagpal2024novel}\\ 
\hline 

\textbf{User and Device Data Privacy} & 51 & \cite{shen2019privacy, liu2018epic, qiao2019effective, zhang2019new, zhang2019provable, qin2019privacy, shin2020privacy, wan2020internet, al2019fog, telikani2023edge, sezer2023ppfchain, sendhil2022verifiable, alshahrani2019secure, sun2022secure, li2022data, li2023three, tao2018multi, sasikumar2023blockchain, deebak2022robust, tudor2020bes, trivedi2023homomorphic, al2019misty, jiang2023federated, shahzad2021privacy, xia2025privacy, agbaje2023privacy, khan2023blockchain, almansour2025privacy, xiao2024efficient, goyal2024enhancing, gamundani2023scalable, wang2024blockchain, nath2024lbpv, li2025distributed, akgun2023privacy, hou2025efficient, jibril2024semantic, 2025rehan, jaganraja2025agile, li2024contract, sahu2024enhanced, mun2024privacy, hindistan2023hybrid, wang2025p3fl, dewangan2023privacy, xia2024decentralized, parmar2023privacy, liu2023ptap, jin2024epaka, wang2025privacy}\\ 
\hline 

\textbf{Location Privacy} & 45 & \cite{long2014achieving, huang2016software, yang2018density, yan2019location, yan2019comprehensive, albouq2020double, zhang2021sport, al2022counterfeit, zhang2022dprl, casanova2023maximizing, liang2020fog, xiong2019locally, an2020lopro, hussain2022improving, kong2019privacy, xia2021effective, fenelon2023private, hussien2020msclp, guo2024dbcpca, de2024protect, zhang2024privacy, verma2024dynamic, wang2025achieving, mukamanzi2022position, enayati2023location, li2025blockchain, xu2024lb, li2025lpps, ma2025personalized, dong2025dummy, bahbouh2024double, qiao2025multi, tan2024privacy, liu2024deep, fu2025prise, han2024cppa, shen2024combining, ameur2024enhancing, wang2024lpf, de2024protect, masood2025dllpm}\\  
\hline 

\textbf{General Privacy} & 99 & \cite{cui2018achieving, sanchez2018integration, yan2019privacy, chen2019internet, lv2019iot, kwabena2019mscryptonet, rahman2019blockchain, baccour2021rl, altowaijri2021privacy, pistono2021cryptosystem, zhou2021secure, li2020deep, xu2019edge, nkenyereye2019towards, deebak2020smart, bera2020designing, gope2018lightweight, gheisari2019context, yang2025pecd, zhao2025blockchain, kumar2025connected, zhao2025blockchain, tang2023efficient, alamer2023privacy, wang2025advancing, dhavamani2024differential, antwi2024enhanced, Jaganraja2025enhancing, yi2024private, liang2025privacy, jain2023design, zhang2024towards, anitha2023unified, suo2023multitiered, mahmood2025privacy, alla2023federated, wang2023privacy, sharma2025blockchain, zhang2025edge, tian2023lightweight, li2025construction, gan2025privacy, sridharan2025lie, zheng2023smart, cao2025privacy, telikani2023edge, chen2023federated, wang2023federated, ranjan2024apps, regan2024balancing, singla2024privacy, abdel2024privacy, mahajan2025deep, kasula2025federated, alrayes2025privacy, gheisari2023agile, karthikeyan2024privacy, li2025aeppfl, xu2025privacy, yenugula2025privacy, hua2024international, singh2026privacy, maniveena2024security1, Al-Balasm2024ehenhancing, sun2024low, liu2024multi, yadav2024hybrid, kil2024optimization, shnain2024privacy, liu2024distributed, jeyakumar2024innovative, Bharat2025fed, gheisari2025enhancing, muntather2025federated, cao2025hybrid, zheng2025awe, zhang2025privacy, li2025aldp, li2024location, mi2024privacy, liu2024privacy, wang2023conditional, yang2023scalable, yazdinejad2024hybrid}\\ 
\hline
\end{tabularx}

\end{table*}

\subsection{RQ5: What privacy types are the main focus of most
PPTs in IoT systems?} According to our review, the three main privacy types in IoT are user, device, and location privacy. Some studies proposed that PPTs can protect the privacy of both the user and the device, as seen in Table \ref{table:Privacy Types}. However, our observation shows that most privacy implementations in IoT devices primarily focus on protecting user data stored in these devices, with very limited or no emphasis on preserving the identity of the IoT device that stores this data. This observation indicates that user privacy is the primary objective of many research studies on PPT in IoT systems.

\subsection{Relationship between Key concepts in IoT system privacy}

In this scoping review, we proposed a privacy preservation taxonomy that includes major privacy entities, as explained in Subsection \ref{privacy-taxonomy}. This taxonomy encompasses privacy-enhancing technologies (PET), privacy preservation techniques (PPT), Privacy Threats, Privacy Goals, and the technologies used to develop PET. Although previous studies have mentioned these entities, none have explored the correlations among them in the context of privacy preservation, particularly within an IoT environment. \autoref{fig:PP-in-IoT} illustrates the relationships among these privacy entities and detailed examples for each entity.
    


\section{Future Direction}
\label{section:future}
After reviewing the PPTs proposed in the selected papers from top literature databases, we identified some future research areas that require the attention of researchers.

    \subsection {Need for device privacy preservation (Device identity protection)} 
    Finn et al. \cite{finn2013seven} and Liyanage et al. \cite{liyanage2018comprehensive} have proposed several types of privacy. However, in a typical user service, provider interactions within an IoT system use two major types of information flow: (a) user data, such as personally identifiable information (PII), including contacts, names, photos, and age, and (b) IoT device data and activity, such as IP address, geolocation, and device ID. These two major data are essential; however, protecting the identity of the device is crucial, as its protection could directly impact the privacy of the user and the location \cite{al2019misty} \cite{al2020overview}. It is essential to note that a compromised IoT device will holistically impact the user and device, as the IoT device itself stores, processes, and transmits these data.

    \textbf{ Research Gap} 
    Despite limited studies on device data privacy, most studies on device privacy still centered around protecting user data stored, processed, or transmitted across these devices in IoT. Therefore, research on device identity privacy is essential for protecting sensitive information about devices in IoT systems. We will be considering both the user and device data privacy in subsequent chapters in this thesis. 
    
    
    \subsection{Need for proper understanding of privacy threats activities} 

    While providing PPTs to mitigate IoT privacy threats is essential, understanding the underlying activities and mechanisms of these threats through systematic privacy threat analysis (PTA) is equally critical. Such analysis provides the foundational insights necessary to design and implement effective mitigation strategies that address the root causes of privacy vulnerabilities.

   \textbf{ Research Gap} 

  Our scoping review revealed a significant gap: most reviewed papers proposing PPTs did not conduct prior analysis of the specific privacy threats they aimed to mitigate. This analytical deficiency likely explains why many reviewed PPTs papers fail to specify the targeted computing layer or privacy type explicitly in their work. A comprehensive PTA systematically identifies vulnerable areas within IoT architectures where privacy threats are most likely to occur, thereby enabling targeted and effective PPT deployment in these critical locations.

  Implementing comprehensive PTA for IoT systems is essential to address the full spectrum of privacy threats, particularly by characterizing threat actors and their behavioral patterns. Machine learning approaches offer a promising avenue for automating threat analysis by training models on privacy-preserving datasets. However, we identified a lack of privacy-focused datasets for training and testing against known privacy threats in IoT systems. To address this methodological gap, a multi-faceted approach can be proposed by generating a synthetic dataset that can simulate realistic threat behaviors in IoT environments, facilitating the development and validation of robust privacy protection measures.  Alternatively, expert elicitation through structured user studies with privacy specialists can provide qualitative insights into emerging threats, attack tactics, and evasion techniques, thereby enriching our understanding of the evolving threat landscape.
  
    \subsection{Importance of Privacy Engineering (PE), Privacy Threat Analysis (PTA), and Privacy Impact Assessment (PIA) in privacy preservation} 
    Privacy measures such as PPT are typically implemented once, during design or during IoT system setup. However, privacy measures should extend beyond the design stage, as many vulnerabilities can arise when the system is operational and are exploitable by a privacy threat. Therefore, the privacy threat mitigation approach should be periodically reviewed, particularly when changes are made to the IoT system, thereby engineering the process of privacy preservation.
    
   Privacy engineering (PE) represents the continuous and comprehensive implementation of privacy protection throughout the lifecycle of IoT systems, extending beyond the initial design phase to encompass operational deployment and maintenance. Within the PE framework proposed by \cite{LINDDUN}, privacy threat analysis (PTA) systematically identifies and analyzes privacy threats, while privacy impact assessment (PIA) evaluates their associated risk levels. These two privacy approaches (PTA and PIA) are fundamental pillars of PE implementation in any operational system.  However, an integrated approach that incorporates these privacy frameworks is essential not only for implementing proactive privacy measures but also for maintaining adaptive, up-to-date privacy protection in IoT systems.

    \textbf{ Research Gap} 
    Integrating privacy measures is crucial in the early stages of an IoT system; however, these measures should extend beyond the design stage. Privacy measures should be incorporated during installation and configuration, and maintained throughout the production use of the IoT system. However, researchers have not effectively emphasized how privacy measures can be implemented continuously beyond the design stage. Introduction of the PE approach, through PTA and PIA, would ensure that privacy is implemented at every stage of the IoT lifecycle, thereby keeping privacy preservation up to date in the IoT system. In the event of any change to the IoT system, whether by adding a new device or modifying configuration, the PTA and PIA should be repeated to ensure that privacy preservation remains up to date. The result will enable prioritization of privacy threats based on their potential risk in IoT, allowing recommendations for possible PPTs to mitigate these threats. Applying this to IoT applications such as Smart Home system should enable proactive and in-depth privacy protection.

     \subsection{The influence of emerging technology}
     Evidence from our scoping review indicates that the introduction of emerging technologies, such as post-quantum cryptography, has contributed to privacy preservation in IoT systems by improving privacy protection \cite{cook2023security, meden2021privacy}. On the other hand, it also provides malicious actors with advanced tools to carry out privacy threats \cite{gupta2023chatgpt, chen2024generative}. This duality highlights the complex balance between utilizing technological advancements for privacy protection and managing the increased risks presented by the sophisticated tools employed by malicious individuals.

    \textbf{ Research Gap} 
   Researchers should leverage emerging technology such as post-quantum algorithms and Large Language Model (LLM) to enhance privacy protection in IoT systems. For instance, integrating an appropriate, lightweight post-quantum cryptography algorithm, with the assistance of generative AI, will provide quantum-safe privacy in an IoT system.

\section{Conclusion}
\label{section:Conclusion}
This scoping review employs the PRISMA-ScR guidelines and Arksey et al.'s framework to comprehensively review PPTs in IoT systems. 

Out of 1177 papers collected, 329 met the selection criteria. These selected papers were analyzed for insights into PPT goals, PET building technologies, PPT implementation in IoT applications, where these PPTs are implemented in IoT computing architectures, and protected privacy types.

 Our review revealed that cryptography is the primary technology used to maintain privacy in IoT systems. At the same time, anonymity remains the primary objective of PPT in IoT implementations, addressing our research questions RQ1 and RQ2, respectively. Furthermore, the continuous expansion of IoT applications across diverse domains, with new introductions in social networks as SIoT and multimedia networks as IoMT, reflects the growing influence of IoT, which answers our RQ3 research question. This study also reveals that PPT is implemented across various IoT computing layers, including cloud, fog, edge, and local networks. However, most PPT implementation occurs at the cloud layer of the IoT system. In addition, medical IoT has more publications on the implementation of PPT than other IoT applications, as our study has gathered, addressing our research question RQ4. We discovered that while most PPT implementations prioritize user privacy, fewer studies focus on device and location privacy, thus answering RQ5. 
 
 Finally, implementing effective PPT to address privacy threats in IoT systems can be challenging due to IoT system resource constraints commonly associated with these systems. Nevertheless, this study highlights our findings through a comprehensive scoping review, outlines future research directions for PPT implementation in IoT systems, and acknowledges the limitations of this work. This review aims to serve as a valuable foundation for researchers in selecting appropriate PPTs for IoT applications and to provide an understanding of the types of privacy these PPTs protect.

\textbf{Limitations}: The study focused on articles published from 2010 to early 2025, potentially missing relevant studies outside this timeframe. The inclusion criteria limited articles to English-language publications, which may have excluded significant non-English research. Additionally, as a scoping review, the study provided a broad overview rather than an in-depth analysis, which may have potentially omitted essential nuances. These limitations should be considered when interpreting the findings.

\section*{Acknowledgement}
This work was partially supported by the Natural Sciences
and Engineering Research Council of Canada (NSERC)
through the NSERC Discovery Grant program.

 \bibliographystyle{elsarticle-num} 
 \bibliography{reference}

\end{document}